\documentclass[twocolumn]{aastex63}
\usepackage{amsmath}

\usepackage{graphicx}
\usepackage{amsmath}
\usepackage{mathtools}
\usepackage{kantlipsum}
\usepackage{xparse}
\usepackage{enumerate}
\usepackage[version=3]{mhchem}
\usepackage{amsmath,array}
\usepackage[T1]{fontenc}
\usepackage{booktabs}

\usepackage{sidecap}
\usepackage{tabularx}
\usepackage{booktabs} % For \toprule, \midrule, and \bottomrule

\received{Received}
\revised{Revised}
\accepted{Accepted}
\submitjournal{ApJ}

\graphicspath{{./}{figures/}}

\begin{document}

\title{A Direct Comparison between the use of Double Gray and Multiwavelength Radiative Transfer in a General Circulation Model with and without Radiatively Active Clouds}

\correspondingauthor{Isaac Malsky}
\email{imalsky@umich.edu}

\author[0000-0003-0217-3880]{Isaac Malsky}
\affil{Department of Astronomy and Astrophysics, University of Michigan, Ann Arbor, MI, 48109, USA}

\author[0000-0003-3963-9672]{Emily Rauscher}
\affil{Department of Astronomy and Astrophysics, University of Michigan, Ann Arbor, MI, 48109, USA}

\author[0000-0001-8206-2165]{Michael T. Roman}
\affil{School of Physics and Astronomy, University of Leicester, University Road, Leicester LE1 7RH, UK}

\author[0000-0002-3052-7116]{Elspeth K. H. Lee}
\affil{Center for Space and Habitability, University of Bern, Gesellschaftsstrasse 6, CH-3012 Bern, Switzerland}

\author[0000-0002-6980-052X]{Hayley Beltz}
\affil{Department of Astronomy and Astrophysics, University of Michigan, Ann Arbor, MI, 48109, USA}

\author[0000-0002-2454-768X]{Arjun Savel}
\affil{Department of Astronomy, University of Maryland, College Park, MD 20742, USA} 

\author[0000-0002-1337-9051]{Eliza M.-R.\ Kempton}
\affil{Department of Astronomy, University of Maryland, College Park, MD 20742, USA} 

\author[0000-0002-6311-4860]{L. Cinque}
\affil{Department of Astronomy and Astrophysics, University of Michigan, Ann Arbor, MI, 48109, USA}

\begin{abstract}
Inhomogeneous cloud formation and wavelength-dependent phenomena are expected to shape hot Jupiter atmospheres.
We present a General Circulation Model (GCM) with multiwavelength ``picket fence'' radiative transfer and radiatively active, temperature dependent clouds, and compare the results to a double gray routine.
The double gray method inherently fails to model polychromatic effects in hot Jupiter atmospheres, while picket fence captures these non-gray aspects and performs well compared to fully wavelength-dependent methods.
We compare both methods with radiatively active clouds and cloud-free models, assessing the limitations of the double gray method.
Although there are broad similarities, the picket fence models have larger day-night side temperature differences, non-isothermal upper atmospheres, and multiwavelength effects in the presence of radiatively active clouds.
We model the well-known hot Jupiters HD 189733 b and HD 209458 b. For the hotter HD 209458 b, the picket fence method prevents clouds from thermostating dayside temperatures, resulting in hotter upper atmospheres and the dissipation of dayside clouds.
Differences in the temperature structures are then associated with nuanced differences in the circulation patterns and clouds.
Models of the cooler HD 189733 b have global cloud coverage, regardless of radiative transfer scheme, whereas there are larger differences in the models of HD 209458 b, particularly in the extent of the partial cloud coverage on its dayside.
This results in minor changes to the thermal and reflected light phase curves of HD 189733 b, but more significant differences for the picket fence and double gray versions of HD 209458 b.
\end{abstract}

\section{Introduction}\label{sec:Introduction}
Although more than 5000 exoplanets have been discovered\footnote{\texttt{https://exoplanetarchive.ipac.caltech.edu/}} \citep{ExoDatabase}, each observation gives us limited data about the physical nature of these alien worlds. Numerical simulations play an important role in augmenting these observations both by aiding interpretation of the data, and guiding the next generation of surveys \citep[e.g.,][]{Heng2015, Madhusudhan2019}. In particular, 3D General Circulation Models (GCMs) are necessary for understanding the inherently multi-dimensional properties of exoplanets and are crucial for comprehensively interpreting these data in observations. However, even the most complex of models must use simplifying assumptions. Creating a computational model that balances the inclusion of complicated physical processes while retaining computational feasibility is a critical part of exoplanet astronomy. Furthermore, less complex models let us isolate how particular physical effects or conditions influence the overall structure of exoplanets and their atmospheres.

The relative ease of observing hot Jupiters (compared to smaller and colder exoplanets) and the lack of a solar system analogue has made hot Jupiters ideal laboratories for atmospheric characterization and expanding atmospheric physics into new regimes \citep{Fortney2021}. Under the standard assumption that these planets have been tidally locked into synchronous rotation states \citep{Rasio1996}, a 3D understanding is of first-order importance, with GCM simulations playing a particularly important role for hot Jupiters. Driven by the need for an efficient and sophisticated model, many groups have developed GCMs \citep[e.g.,][]{Showman2002, Showman2009, Heng2011, Rauscher2012, Dobbs2013, Mayne2014, Cho2015, Heng2015} to study these extreme planets. These models show many common features: transonic winds with an eastward equatorial jet on the order of several kilometers per second, a hot-spot that has been advected east of the planet's substellar point, and day-night temperature differences exceeding several hundred Kelvin. Furthermore, the diversity of assumptions inherent in these GCMs and the universality of these resulting features solidifies the field's confidence in the overall results, at least for planets with clear atmospheres and no additional physical processes. 

\textcolor{black}{However, there are many details of the circulation patterns that are subject to the particulars of modeling approaches. This includes physical effects (as discussed below), but even the numerical choices within a GCM can influence the predicted atmospheric structure. One issue in particular is that of the treatment of numerical dissipation, which must be included in order to prevent a buildup of small scale noise, which should physically be dissipated at sub-grid scales. Work by \citet{Thrastarson2011} discussed numerical dissipation in the context of hot Jupiter atmospheres, pointing out that there is no \textit{a priori} way to determine the correct strength of numerical dissipation to use, but rather multiple values should be tested to decide the appropriate level and, importantly, the strength of dissipation should be expected to change with the atmospheric conditions modeled as well as the spatial resolution. Of further concern, the strength of numerical dissipation used can change the overall wind speeds in a model \citep{Heng2011a} and if the dissipated kinetic energy is not returned as thermal heating, this will violate energy conservation within the model and could change the nature of the circulation, especially if the dissipation is strongly localized \citep{Goodman2009,Rauscher2012_energetics}. Then there is also the question of the nature of the dissipation; the modeled flows are often transonic and shocks or turbulence triggered by instabilities could shape the circulation \citep{Li2010}. It is a challenge to numerically model these sources of dissipation, as they are on scales smaller than typically included in full GCMs, but are dependent on the large scale flow conditions that are driven globally. There continues to be ongoing work exploring various forms of dissipation and the roles they may or may not play in shaping hot Jupiter circulation \citep{Dobbs2013,Fromang2016,Ryu2018,Menou2020,Menou2022}.}

\textcolor{black}{In addition to the accurate treatment of various sources of sub-grid dissipation, there are additional physical effects that we expect to shape hot Jupiter atmospheres.} Magnetic drag, disequilibrium chemistry, inhomogeneous cloud formation, and multiwavelength radiative transfer all impact the thermal structure of the planet and create feedback effects \citep[e.g.,][]{Showman2009, Rogers2014, Kataria2016, Lee2016, Parmentier2016, Helling2016, Amundsen2016, Drummond2018, Mendon2018a, Mendon2018b, Tan2019, Steinrueck2019, Roman2019, Steinrueck2021, Beltz2022, Deitrick2022}. To our knowledge, there currently exists no hot Jupiter model capable of simulating in concert all of the physical processes listed above. A single model that includes all of the important physical processes---even simplified mechanisms---would be an important addition to the field. However, including all relevant effects is not as simple as iterating on existing GCMs, as each new addition increases the computational expense of simulations. Therefore, it is critical to determine which simplifying assumptions are the most physically valid and how accurate models can be created with low computational overhead.

In this paper we focus on the interaction between active, temperature-dependent clouds and the radiative transfer scheme used in our GCM. Previous cloud modeling with our GCM used a simplified, double gray radiative transfer scheme, with possible limitations including the excess depression of thermal emission from the cloudy nightside of the planet \citep{Roman2017,Roman2019,Roman2021}. Here, we present a new, more complex, picket fence radiative transfer routine, which can more accurately capture the interaction of the radiation field with the clouds, in a computationally efficient way.

The radiative transfer scheme in a GCM can be implemented at varying levels of sophistication and computational expense. In less complex realizations, gaseous absorption coefficients are chosen for two bands \citep{Guillot2010, Hansen2010, Rauscher2012}. This is known as the semi-gray or double gray radiative transfer routine\footnote{Of note, this is separate from two-stream radiative transfer. Two-stream refers to the the radiative transfer approximation where radiation is integrated over two discrete directions (up and down), while the double gray terminology refers to the choice to use two absorption coefficients to calculate optical depths.}. At least two bands are necessary in order to correctly treat the incident stellar irradiation (which is attenuated as it descends into the atmosphere) and the thermal radiation from the planet (dependent on the atmospheric temperature structure), as the geometric assumptions vary between these distinct components of the radiative transfer. \textcolor{black}{Inherent in the double gray approach of \cite{Guillot2010} is that the Planck weighted, absorption weighted, and flux weighted opacities are all equal \citep{Heng2014}. This results in no opacity windows for the gas to cool, and artificially isothermal upper atmospheres.}

Although a double gray radiative transfer routine is efficient and can capture overall properties of the expected atmospheric profiles, there are several physical aspects that will not be accurately captured. First, regardless of how one picks the two absorption coefficients, the temperatures near the top of the atmosphere will be systematically hotter than predicted by multiwavelength solutions, as non-gray effects allow the upper atmosphere to cool \citep{pierrehumbert2010, Parmentier2014}. This effect can be several hundred Kelvins for hot Jupiters. Furthermore, double gray models necessarily produce isothermal upper atmospheres, whereas non-gray effects result in more realistic temperature-pressure profiles. Additionally, while clouds blanket outgoing thermal flux in the infrared, at long enough wavelengths clouds should become optically thin and allow emission to space. The double gray treatment, with its single absorption coefficient for the thermally emitting planet layers, cannot capture these behaviors.

Another common radiative transfer implementation uses the \textcolor{black}{k-distribution method with the correlated-k approximation} for calculating atmospheric opacities \citep{Goody1989, Lacis1991, Fu1992}, such as in \cite{Showman2009} (SPARC), \cite{Amundsen2016} (The UK Met Office GCM), or \cite{Lee2021} (Exo-FMS). In the \textcolor{black}{k-distribution} method, opacities are calculated at millions of different wavelengths and then sorted by strength and binned into a prescribed number of wavelength bands. One benefit of the \textcolor{black}{k-distribution} method is that opacities can be pre-calculated for a grid of pressure-temperature points and take into account molecular and atomic line opacities, as well as line broadening. The \textcolor{black}{k-distribution} scheme finds fluxes that usually agree with line-by-line calculations to within a few percent or better  \citep{Showman2009,Amundsen2017}. However, while significantly faster than a line-by-line calculation, the \textcolor{black}{k-distribution} scheme is still computationally expensive due to the need to solve the radiative transfer equations for each opacity bin.

Recently, a non-gray picket fence method has been shown to produce similar results to the \textcolor{black}{k-distribution} radiative transfer schemes in GCMs, but at a fraction of the computational cost \citep{Parmentier2014, Parmentier2015, Lee2021}. In the picket fence method \citep{Chandrasekhar1935, Mihalas1978}, opacities are calculated separately for the thermal and the incident starlight radiation. The thermal calculation has two bands, one which represents the molecular and atomic line opacity and one which represents the general continuum opacity. A variable number of bands can be used to calculate the propagation of radiation in the starlight bands. The picket fence method self-consistently links the opacity of each layer to the local pressure and temperature through pre-calculated fitting functions for a range of planetary effective temperatures \citep{Parmentier2015}. \citet{Lee2021} demonstrated that, in the case of a clear hot Jupiter atmosphere, running a GCM using a radiative transfer routine with picket fence opacities produces results in good agreement with the same GCM using \textcolor{black}{k-distribution} radiative transfer, at much lower computational cost, and was a significant improvement over using a double gray scheme.

A key goal of this work is to understand how well the picket fence radiative approach performs when coupled with cloudy GCMs. Available observational evidence strongly suggests that clouds are pervasive in hot Jupiters, and therefore a key element of hot Jupiter GCMs. They play an important role in shaping both their atmospheric structure and subsequent observables \citep{Demory2013, Shporer2015, Sing2016, Barstow2017, Gao2021}. Cloud coverage is critical for radiative transfer, as clouds interact with the outgoing thermal radiation from the planet atmosphere, as well as the reflecting and scattering the incident stellar radiation. \textcolor{black}{Several groups have created GCMs capable of modeling cloud coverage, using extinction approximations \citep[e.g.,][]{Lee2016}, or explicitly as a source of scattering \citep[e.g.,][]{Lines2019, Roman2019, Christie2022, Lee2023}.} GCMs have shown that radiative feedback from clouds plays an important role in shaping the thermal structure of hot Jupiters, especially at lower irradiation levels \citep{Lee2016, Lines2018, Roman2021}. Although not the focus of this paper, clouds also manifest in transmission spectra and are an important consideration when characterizing these observations \citep[e.g.,][]{Wakeford2015}, especially as transmission spectra are more susceptible to clouds because of slant geometry effects.

This paper is organized as follows: in \S~\ref{sec:Methods} we present our methodology for modeling the double gray and multiwavelength GCMs. These methods are further broken down into sub-sections on the gas radiative transfer, the cloud species and condensation curves we implement, and the cloud scattering and absorption. In \S~\ref{sec:Results}, we present results for the clear and cloudy models of HD 189733 b and HD 209458 b. Finally, a discussion of the results and the conclusions are in \S~\ref{sec:Conclusions}.

\section{Methods}\label{sec:Methods}
In order to compare GCMs simulated with radiatively active clouds and either double gray or multiwavelength radiative transfer schemes, we model two hot Jupiters (HD 189733 b and HD 209458 b, \citealt{Bouchy2005, Charbonneau2000}) for a range of aerosol parameterizations. The GCM presented here (which we refer to as the RM-GCM) has undergone significant iterative extensions from its inception as a meteorological model for Earth using Newtonian relaxation \citep{Hoskins1975}. It has since been adapted for hot Jupiters \citep{Menou2009, Rauscher2010} and updated to include double gray radiative transfer \citep{Rauscher2012,Roman2017}, magnetic drag and localized Ohmic dissipation \citep{Rauscher2013}, inhomogeneous aerosols and radiatively active temperature-dependent clouds \citep{Roman2019}, and clouds with pressure dependent scattering and absorption \citep{Roman2021}. In this work, we present two important additions. First, a new picket fence radiative transfer scheme. Second, an upgraded treatment of radiatively active clouds that is self-consistent with our new radiative transfer scheme.

HD 189733 b and HD 209458 b have been extensively characterized in observational works as well as computational studies. Simulating these two planets allow us to explore different irradiation temperatures and benchmark our results. In particular, we chose HD 189733 b to demonstrate the effects of cloud formation on a colder planet, where we expect cloud formation throughout the planet atmosphere, in contrast with the hotter planet HD 209458 b where clouds are mostly expected just on the cooler nightside. We take most model parameters from \cite{Rauscher2012}, but see Table \ref{tab:planet_parameters} for the full set of planet parameters.

\begin{deluxetable*}{llll}
{\tablehead{\multicolumn{4}{c}{Planet and Star Parameters}}}
\startdata
\multicolumn{1}{c}{Parameter} & \multicolumn{1}{c}{HD 209458 b Value}  & \multicolumn{1}{c}{HD 189733 b Value} & \multicolumn{1}{c}{Units} \\
\hline
\hline
Semi-major Axis                                        & 0.047                  &0.031 & au                \\
Stellar Effective Temperature                          & 6071                   &4875& K                 \\
Stellar Radius                                         & 1.19                   &0.81& R$_\odot$         \\
Gravitational Acceleration                             & 9.0                    &21.2& m s$^{-2}$        \\
Planet Radius                                          & 9.9$\times$10$^7$      &8.0$\times$10$^7$& m                 \\
Internal Temperature (T$_{int}$)                         & 500                  & 365 & K                 \\
Planetary Rotation Rate                                & 2.063$\times$10$^{-5}$ &3.277$\times$10$^{-5}$& radians s$^{-1}$  \\
\hline
\multicolumn{4}{c}{\textit{Double Gray Specific Parameters}} \\
\hline
\hline
Infrared absorption coefficient $\kappa_{\mathrm{IR}}$ & 1.00$\times$10$^{-2}$  &1.00$\times$10$^{-2}$& cm$^2$ g$^{-1}$   \\
Optical absorption coefficient $\kappa_{\mathrm{VIS}}$ & 4.00$\times$10$^{-3}$  &4.00$\times$10$^{-3}$& cm$^2$ g$^{-1}$   \\
IR photospheric pressure ($\tau$=2/3, cloud free)          & 60                     &141& mbar              \\
\hline
\enddata
\label{tab:planet_parameters}
\caption{The parameters used for our models of HD 209458 b and HD 189733 b.}
\end{deluxetable*}

\subsection{Picket Fence Radiative Transfer}
Here we present an updated radiative transfer routine for our GCM. Past versions of this model have used a double gray radiative transfer scheme, which solves the radiative transfer equations for two distinct wavelength bands: one for the thermal emission of the planet and one for the incident starlight. \cite{Roman2017} previously implemented the radiative transfer methods from \cite{Toon1989} to allow for the calculation of inhomogeneous multiple-scattering atmospheres \textcolor{black}{for both starlight and thermal emission. In \cite{Roman2017}, and this work, we use the two-stream hemispheric mean approximation for the scattering source function. Furthermore, for the source function technique we use three Gauss points.} We expand the optical depth calculations, the number of radiative transfer bands, and the cloud parameterization \textcolor{black}{for the RM-GCM. While this picket fence radiative routine provides increased physical accuracy over the double gray version, our two-stream approximation nevertheless still limits the accuracy of scattering in the large-particle limit \citep{Kitzmann2013,Heng2017a}, although this could be addressed in future updates \citep{Heng2018a}.}

The first major change to the GCM is the implementation of the picket fence radiative transfer method from \cite{Parmentier2014}. In order to calculate the gas contribution to the optical depth at each band, we use the fitting functions and analytical results from \cite{Parmentier2015}. Similar to the double gray approach, the picket fence allows us to treat the thermal emission and the starlight radiative transfer independently. First, we increase the number of radiative transfer bands within the GCM from two to five. Of these five, two are for the thermal calculations, and three are for the incident starlight. However, the radiative transfer calculations for the incident starlight can be calculated for any number of bands. We follow \cite{Lee2021} and use three bands for the incident starlight because this was the minimum number (and therefore fastest computationally) that matched the numerical solution to within 1\% accuracy \citep{Parmentier2015}. Of note, these bands do not correspond to specific wavelength ranges, but rather separate bands for the starlight and thermal radiation, and opacities within each chosen to model the net radiative transfer.

For each band, we find the layer optical depths using the fitting coefficient tables of \cite{Parmentier2014} and \cite{Parmentier2015}. To find the thermal band Rosseland mean opacity as a function of pressure, temperature, and atmospheric metallicity we use the \textcolor{black}{solar} \cite{Freedman2014} fitting functions. We also include functionality for the \cite{Valencia2013} opacity fitting function. Both implementations are useful for exploring a range of temperature and pressures, but the \cite{Valencia2013} function can systematically underestimate Rosseland mean opacities \citep{Parmentier2015} so we restrict ourselves to the \cite{Freedman2014} fitting functions. \textcolor{black}{The coefficients and analytic fitting functions used to determine the layer optical depths are specific to solar metallicity atmospheres in \textcolor{black}{local} chemical equilibrium and G-type stars, but can be expanded to include a broader range of exoplanet atmospheres.}

Previous works within the community have referred to the thermal bands as the IR bands and the visible bands as optical bands. However, this notation may be confusing as the optical and IR bands may not align with optical and infrared wavelengths, as noted in \cite{Lee2021}. This extends back to \cite{Toon1989}, where IR and optical bands correspond not to wavelength regimes, but rather to whether the calculated fluxes are from the thermal emission of the planet or are from the incident starlight which, for the case of the Earth and other Solar System planets, are nicely separable into infrared and optical regimes. Throughout this work, we adhere to the convention of calling the two distinct bands the thermal and starlight bands, but point out \textcolor{black}{this source of confusion} in the literature.

To calculate the optical depths for a single planet temperature-pressure profile, we start with the planet irradiation temperature ($T_{irr}=T_{\mathrm{eff,\odot}}\sqrt{R_{\mathrm{\odot}}/a}$) and the planet internal temperature ($T_{int}$). For both HD 189733 b and HD 209458 b we follow the $T_{int}$-flux relation in \cite{Thorngren2019}, who parameterized the observed trend that more irradiated hot Jupiters have larger radii and so higher entropy interior adiabats. \cite{Thorngren2019} showed that while Jupiter has an intrinsic temperature around 100 K, hot Jupiters likely have intrinsic temperatures several hundred Kelvin hotter, even up to 700 K. A hotter intrinsic temperature moves the radiative-convective boundary to lower pressures and can potentially influence the circulation and thermal structure in the observable atmosphere \citep[e.g.,][]{May2021, Komacek2023}. In addition to including $T_{\mathrm{int}}$ in our bottom boundary conditions for the thermal flux in the \cite{Toon1989} radiative transfer calculations, we also include it in calculating the optical depths from fitting coefficients in the picket fence radiative transfer method.

Next, we follow the picket fence routine used in \cite{Lee2021} to calculate the gas opacities at each layer by using the pre-calculated coefficients and analytic functions to determine opacities in each band. Finally, the total layer gas optical depth for each band is calculated assuming hydrostatic equilibrium as

\begin{equation}
\tau_{g, n} = \frac{\kappa(P, T, n)}{g} \Delta P,
\end{equation}

\noindent where $\tau_g$ is the layer gas optical depth, $\kappa$ is the pressure and temperature dependent layer absorption in that band, $g$ is the gravity, and $\Delta P$ is the difference in pressure between the top and bottom of the layer.

There are two contributions to the layer optical depth within the GCM. First, the gas optical depth ($\tau_g$). Second, there is a contribution from each of the cloud species (discussed below) that is calculated separately from the picket fence optical depth. Throughout this work $\tau$ refers to the \textit{layer} optical depth rather than the cumulative optical depth from the top of the atmosphere, and the sum from the gas and cloud contributions.

Simultaneously with the new picket fence gas radiative transfer, we maintain the double gray radiative transfer scheme, keeping the less sophisticated method in order to compare the new results to previous models and for instances where the more simplified scheme may be desired in future works. Because our picket fence radiative transfer scheme sums the heating contributions from each of its five bands linearly, the double gray results can be retrieved by setting all three starlight band absorption coefficients equal and the two thermal absorption coefficients equal.

Our implementation of the picket fence radiative transfer scheme within the GCM is a significant methodological addition, and also introduces several avenues of expansion for future work. However, in order to present a clear and concise demonstration of these new methods, we restrict our simulated suite to a subset of the different parameterizations made possible with these expansions. For example, the analytical fitting coefficient tables from \cite{Parmentier2015} can also be implemented without opacity contributions from the upper atmosphere absorbers TiO and VO, allowing users to test the effect of having these stratospheric absorbers present. Here we assume solar metallicity and include TiO and VO by default.

Last, we include the effects of Rayleigh scattering by imposing a global Bond albedo at the top of the atmosphere, reducing the incoming stellar flux before it enters the atmosphere (and radiative transfer routine). Physically, Rayleigh scattering is expected to cause radiative scattering and reflect some portion of the incident starlight. Rayleigh scattering also increases the opacity of the atmosphere, reducing the depth that starlight penetrates into the atmosphere, but this effect is not included by \textcolor{black}{our approach of simply imposing a global Bond albedo}. Rayleigh scattering is present for both cloudy as well as clear atmospheres, where it is the only cause of reflected starlight. In this work, we use a constant base Bond albedo of 0.10 for both HD 189733 b and HD 209458 b (cloudy models add an additional source of scattering and have larger Bond albedos). This value was chosen to \textcolor{black}{approximately} match the results of sophisticated 1D models with non-equilibrium chemistry and multiwavelength radiative transfer post-processing, as expanded upon in the Appendix. Because of this choice, we did not use the Bond albedo function from \cite{Parmentier2015}.

\subsection{Cloud Species}
In addition to updating the gas opacities within our radiative transfer scheme, we also take this opportunity to calculate the scattering and absorption from atmospheric aerosols. We follow the implementation of clouds outlined in \cite{Roman2021}. \textcolor{black}{Based on  \cite{Mbarek2016}, \cite{Roman2019}, and \cite{Roman2021}}, we allow for the following species: KCl, ZnS, Na$_2$S, MnS, Cr, SiO$_2$, Mg$_2$SiO$_4$, VO, Ni, Fe, Ca$_2$SiO$_4$, CaTiO$_3$, and Al$_2$O$_3$, as they have condensation curves within the pressure-temperature regime of hot Jupiter atmospheres \citep{Mbarek2016}. To calculate the mole \textcolor{black}{(number)} fractions of each species, we first determined the abundance of the limiting atoms for each species assuming a solar elemental abundance.\footnote{In order, the limiting atoms were K, Zn, Na, Mn, Cr, Si, Mg, V, Ni, Fe, Ca, Ti, and Al.} The atmospheric abundances for each species were then calculated using the data from \cite{Burrows1999} and \cite{Anders1989}. \textcolor{black}{Although clouds play an important role in hot Jupiter atmospheres, the exact cloud species expected to form are uncertain \citep[e.g.,][]{Heng2013, Morley2013, Lee2015}. While distinguishing degenerate cloud compositions is spectroscopically challenging, there are some notable observational tests that may be possible to differentiate silicate vs. iron atmospheres, especially with new capabilities on JWST. It may be possible to distinguish cloud species based on vibrational modes \citep{Wakeford2015}, through a combination of albedo and phase curve measurements \citep{Heng2013}, or through absorption and scattering properties \citep[e.g.,][]{kok2011}.}

One change from the cloud implementation in \cite{Roman2021} is our substitution of Cr$_2$O$_3$ for Cr. This substitution was made due to the lack of available scattering data for Cr$_2$O$_3$, which were simply assumed based on other oxides. However, there is significant uncertainty in which oxide forms \citep{Roman2021}, and other works have predicted Cr in hot Jupiter atmospheres \citep[e.g.,][]{Powell2019}, making our substitution reasonable. Last, recent work has suggested that including all of these species above may overestimate the cloud coverage on hot Jupiters due to the low nucleation rates of some species \citep{Gao2020}. Therefore, our default cloud assumption does not include ZnS, Na$_2$S, MnS, Ni or Fe, but we retain the option of allowing these additional species to test how they can shape the atmosphere and observable properties (especially the highly absorptive Fe), and to compare to previous work \citep{Roman2021}.

\subsection{Condensing Active Clouds}
\textcolor{black}{We model scattering and absorbing clouds in the RM-GCM, using a similar treatment to \cite{Roman2019} with several updates pertaining to the number of radiative transfer channels and the wavelengths at which the cloud radiative properties are evaluated.} We use the condensation curves from \cite{Mbarek2016}, and supplemented with the Cr condensation curve from \cite{Morley2012}. Clouds only condense at locations where the local temperature profile drops below the condensation temperature of each species.

Similar to \cite{Roman2019}, since we do not directly predict the balance between vertical mixing of cloud particles and their size-dependent gravitational settling, we parameterize the vertical extent of the clouds as either ``extended" or ``compact" within the GCM. For the extended cloud models, clouds form at all layers where the temperature is lower than the condensation curve for each species. In contrast, for the compact cloud models the base of the cloud layer forms where the vertical temperature profile first drops below the condensation curve (going from higher to lower pressure) and then we truncate the clouds after 5 vertical layers above that, which corresponds to approximately 1.4 scale heights for our models. For both the compact and the extended cloud models, we allow clouds to potentially exist up to 10$^{-5}$ bar, the lowest pressure included in our models.

These two options, compact or extended cloud distributions, provide bounding examples for whether we expect larger particles and weak vertical mixing (compact clouds) or small particles and strong vertical mixing (extended clouds). There is currently uncertainty as to which outcome is most likely, with some works favoring relatively compact clouds \citep[e.g.,][]{Ackerman2001} and some dynamical models suggesting more extended clouds \citep[e.g.,][]{Parmentier2013, Lines2018}. \textcolor{black}{Vertical cloud extent is shaped by complex, multi-scale processes, as discussed in more detail by other works \citep[e.g.,][]{Parmentier2013, Lee2015, Helling2016, Roman2019, Komacek2019, Komacek2022}.Here we choose a simple, explainable parameterization it in lieu of complex modeling.}

Where clouds are present, the mass of condensate is calculated at each pressure as in \citet{Roman2017}:
\begin{equation}
    m_{c,i}(p) = f \frac{\chi_i \mu_i}{\overline{\mu}}\frac{\Delta P}{g}
\end{equation}
\noindent where $m_c$ is the condensate mass per unit area, $\chi_i$ is the mole fraction, and $\mu_i$ the molecular weight of species $i$, while $\Delta P$ is the pressure difference between model layers, and  $\overline{\mu}$ is the mean molecular weight of the atmosphere. Following \cite{Roman2021} throughout this work we assume that the fraction of each species that condenses, $f$, is equal to 10\%. However, in some cases this level of condensation was \textcolor{black}{numerically} unstable \textcolor{black}{due to rapid changes in layer optical depths (as noted in \cite{Roman2021})} and we reduced the condensation fraction to allow the models to run for 1000 days.\footnote{These parameters are shown for each model in the Appendix, Table \ref{tab:all_model_parameters}}. Following \cite{Roman2021}, we use a pressure dependent vertical gradient for particle sizes: for pressures less than 10 mbar the radius is fixed at 0.1 $\mu$m, then increasing exponentially in size, reaching almost 80 $\mu$m at 100 bar (the bottom boundary of our model). \textcolor{black}{For each particle size, we calculate the absorption, scattering, and extinction parameters assuming a mono-disperse population and spherical particles following using the methods in \cite{Kitzmann2018}.}

The clouds within our GCM are radiatively active and shape the atmospheric temperature structure,  with feedback on the cloud distributions. For each cloud species we perform the condensation and radiative transfer calculations for each radiative timestep. This treatment is in contrast to simpler post-processing work, where GCMs are run without clouds and then clouds are added in after run-time based on the final temperature structure of the planet. Post-processed clouds are computationally efficient, but lack the ability to self consistently model cloud formation and the resulting effects on the atmospheric radiative transfer. Furthermore, radiative feedback has been shown to have a non-negligible impact on the atmospheric thermal structures of hot Jupiter atmospheres \citep{Roman2019}.

\textcolor{black}{Implicit within our cloud scheme is the assumption that the cloud formation and evaporation timescales are less than the atmospheric flow timescale, as our clouds form and dissipate instantaneously with temperature changes at each radiative timestep. \cite{Powell2018, Powell2019} simulate the timescales of various cloud microphysical processes and find the evaporation timescale to be effectively instantaneous compared to the dynamical timescales (several seconds compared to the $\gtrsim$ hundreds of seconds for advection from one grid point to the next). However, the growth timescales, especially at pressures less than 10$^{-2}$ bar, are comparable to the hemisphere advection crossing time. This means that we may overestimate nightside clouds, especially just past the evening terminator, since it may take some time for the clouds to grow once the gas blown over from the dayside cools.}

\subsection{Cloud Radiative Properties}
The second major change to the RM-GCM in this work is our implementation of a more robust cloud parameterization, and one that is consistent with the picket fence gas optical depth calculations. We expand our cloud treatment to up to five bands, matching the radiative transfer scheme described previously. We chose fixed wavelengths to evaluate both the reflected starlight cloud scattering and absorption properties for the picket fence models at 500, 650, and 800 nm. These wavelengths were chosen to span the dominant wavelength range of the incident radiation from Sun-like host stars. The cloud radiative transfer properties for the double gray models were evaluated at a single wavelength. For all starlight channels we chose 500 nm.

The polychromatic clouds represent a significant improvement in the cloud scattering and absorption calculations from \cite{Roman2021}, as the three starlight bands more accurately capture the radiative transfer effects than the single starlight wavelength band used previously. Furthermore, for the two thermal bands we use temperature and pressure dependent Rosseland and Planck mean weighted cloud properties rather than values at any single wavelength. We chose the Rosseland and Planck mean values to best capture how the weighted extinction opacities (as a function of wavelength) dominate how radiation propagates through the clouds.

Each cloud species can affect the radiative transfer calculations through its absorption and scattering properties. Following \cite{Roman2021}, the cloud optical depth of species $i$ at each pressure is
\begin{equation}
    \tau_{c,i} = \frac{3 m_{c,i}(p) Q_{ext,i} (p)}{4 r(p) \rho_i}
\end{equation}
\noindent where $r(p)$ is the particle size and $\rho_i$ is the particle density. We use Mie theory code based on \cite{Wolf2004} to pre-calculate tables of scattering and absorption parameters for the extinction coefficient (Q$_{ext}$), single scattering albedo ($\varpi_0$), and asymmetry parameter ($g_0$). For each cloud species we calculate the scattering and absorption properties from the wavelength dependent refractive indexes ($n$) and extinction coefficients ($k$) from \cite{Palik1997} and \cite{Kitzmann2018}. Furthermore, when calculating the Rosseland and Planck mean parameters for the thermal bands, we follow \cite{Lef2022} and calculate the pressure and temperature dependent Rosseland and Planck mean parameters. All optical property calculations are done over a grid that covers particle sizes from 1$\times$10$^{-7}$~m to 1$\times$10$^{-4}$~m and temperatures from 500 K to 3000 K. These grids span the range of temperatures and particle sizes expected in our GCM. Further work with ultra-hot Jupiters may necessitate expanding the temperature grids past 3000 K.

For each of the thirteen possible cloud species, we calculate Q$_{ext}$, $\varpi_0$, and $g_0$. When the GCM runs, during each radiative timestep and for each layer, the code checks whether a cloud condenses and if so, interpolates within the pre-computed scattering and absorption data tables for the appropriate cloud species to find the pressure dependent values (for the starlight bands) or the pressure and temperature dependent values (for the thermal bands).

We can recreate the double gray clouds scheme by calculating scattering and absorption properties at 500 nm and 5000 nm. This allows us to be fully consistent when comparing the differences between the double gray radiative transfer calculation and the new picket fence calculations. For the double gray radiative transfer we adjust the $\beta$ values (the fraction of total flux that passes through each radiative transfer band) to simplify to one starlight band and one thermal band. For more details about $\beta$, see \cite{Parmentier2014}. This simple addition of the Rosseland and Planck mean cloud optical depths with the gas opacities is not strictly physical \citep{Heng2017}, but represents the cloud feedback mechanisms well enough within the picket fence radiative transfer scheme.

\section{Results}\label{sec:Results}
\begin{figure*}[!htb]
\begin{center}
\includegraphics[width=0.6\linewidth]{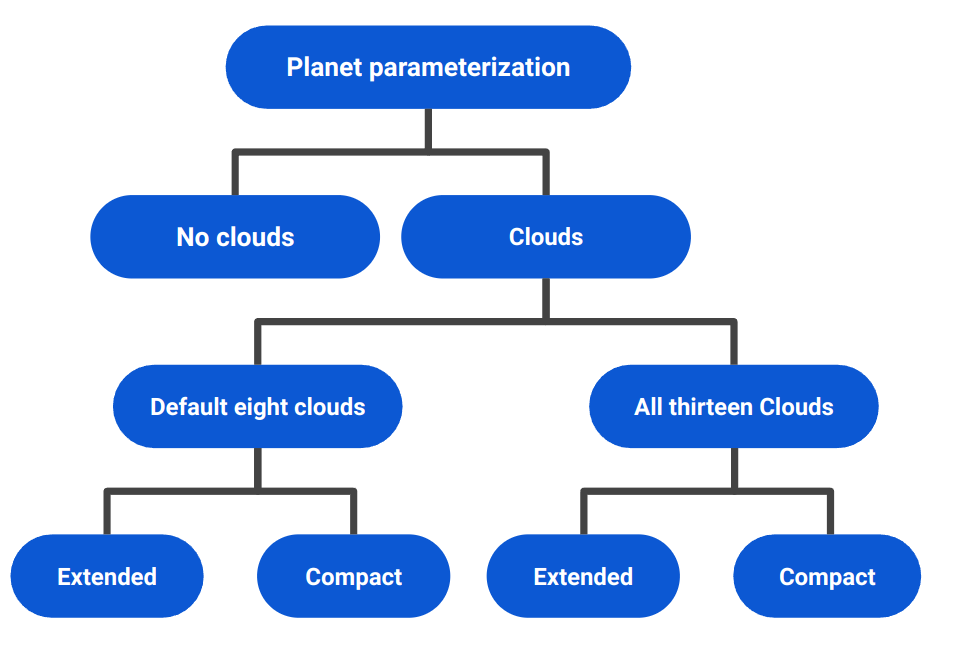}
\caption{A tree diagram showing how we parameterized the different cloudy or cloud free distributions with each GCM. For each planet we ran two sets of these models, one using double gray radiative transfer and one with our new picket fence scheme.}
\label{fig:model-flowchart}
\end{center}
\end{figure*}

We ran a suite of 20 models of HD 189733 b and HD 209458 b: double-gray and picket fence versions of each planet, with the five different cloud parameterizations as shown in Figure \ref{fig:model-flowchart}. All models also were run at a horizontal spectral resolution of T31 (comparable to 48 latitudes points and 96 longitudes points) and 50 pressure levels evenly distributed in log space from 10$^{-5}$ to 10$^{2}$ bar. Initial pressure-temperature profiles at all locations were set using the relations in \cite{Guillot2010}, using the double gray absorption coefficients in Table \ref{tab:planet_parameters}. We assumed an $f$ value of 0.25 following \cite{Burrows2003} to average flux over the entire planet surface\textcolor{black}{, and a hyper-diffusion timescale of 0.0025 planetary days}. All models were run for 1000 planet days. More details about the model parameterizations can be found in Table \ref{tab:all_model_parameters}.

There are significant differences between the upper atmospheres of the two radiative transfer schemes in terms of wind patterns, temperature structure, and cloud distributions. Our picket fence models no longer tend towards the isothermal upper atmospheres characteristic of the double gray method. Instead, they show dayside temperature inversions of hundreds of degrees Kelvin, particularly in the case of HD 209458 b. Furthermore, the HD 209458 b picket fence models have day-night temperature differences hundreds of degrees larger than those of the double gray models and different cloud distributions (with the prominent differences due to the hotter day sides of the picket fence models). These differences persist under the presence of clouds, and the multiple radiative transfer bands of the picket fence scheme better captures the inherent polychromatic nature of atmospheric clouds. In contrast to the upper atmospheres, we find similar deep planet atmospheres for the picket fence and double gray radiative transfer schemes.

\subsection{Vertical Atmospheric Profiles}
\begin{figure*}[!htb]
\begin{center}
\includegraphics[width=0.95\linewidth]{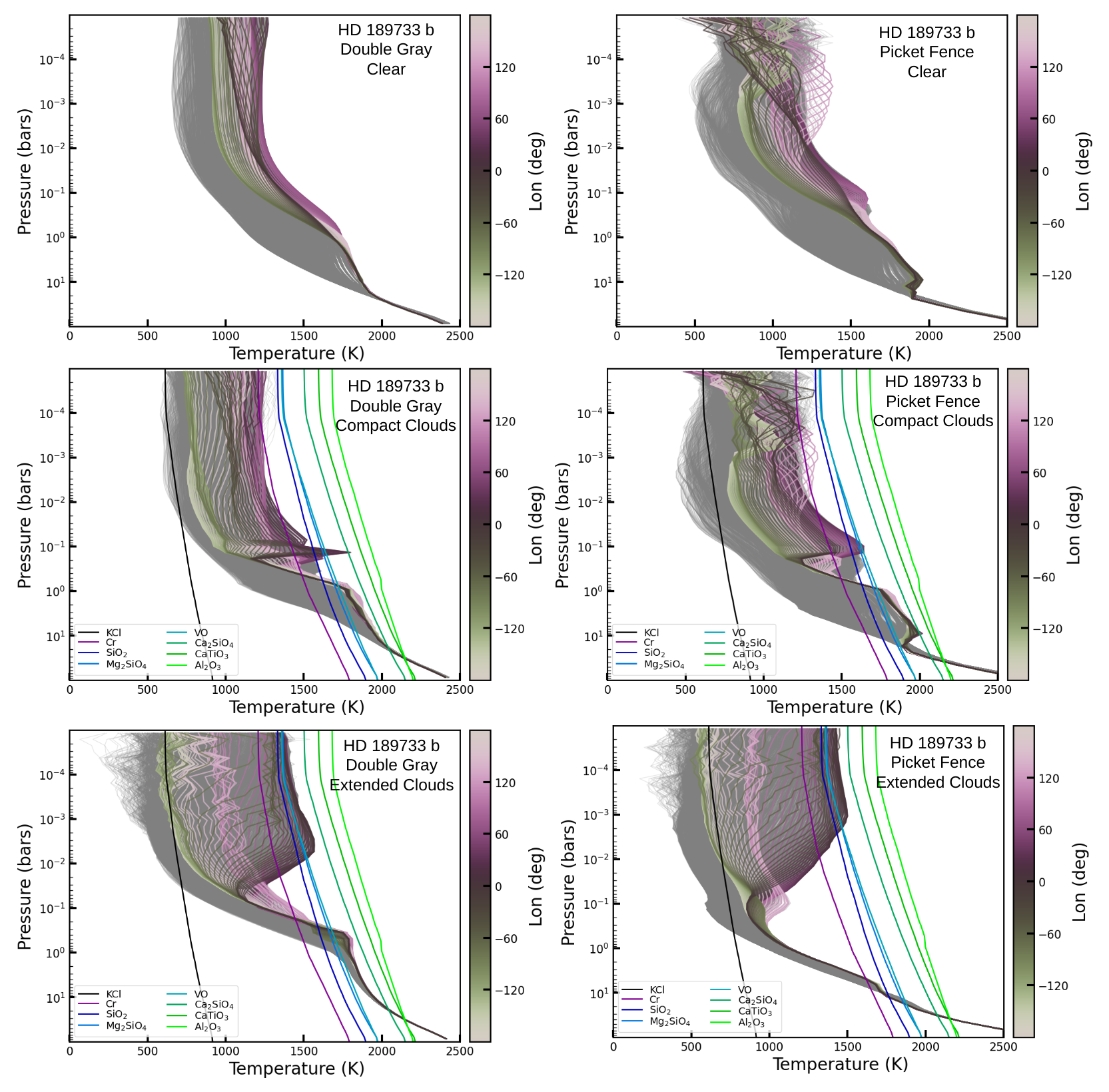}
\caption{Pressure-temperature profiles for HD 189733 b. The colored lines are equatorial profiles, with the color bar denoting the longitude (zero is substellar). Additional colored lines correspond to the condensation curves for the 8 cloud species used for the standard cloudy models. The gray lines show the pressure-temperature profiles for all non-equatorial profiles.}
\label{fig:PT-HD189}
\end{center}
\end{figure*}

\begin{figure*}[!htb]
\begin{center}
\includegraphics[width=0.95\linewidth]{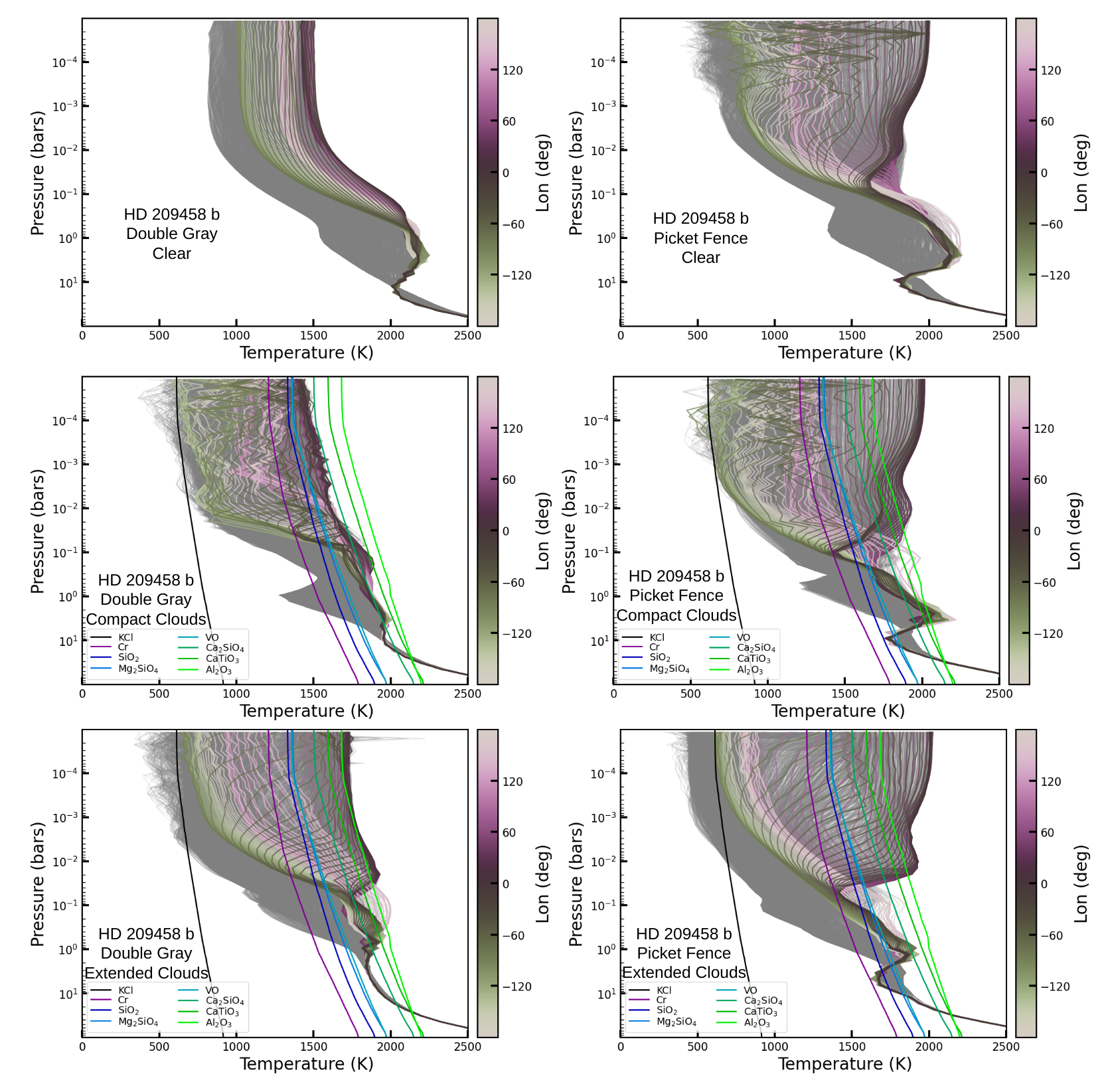}
\caption{Pressure-temperature profiles for HD 209458 b. All details match those described in Figure \ref{fig:Averaged-Clouds-HD189}.}
\label{fig:PT-HD209}
\end{center}
\end{figure*}

The differences between the double gray and picket fence radiative transfer routines are most apparent in the pressure temperature profiles, particularly in the upper atmospheres. Figures \ref{fig:PT-HD189} and \ref{fig:PT-HD209} show the pressure-temperature profiles of HD 189733 b and HD 209458 b, for versions of these planets without clouds and with the default cloud species, both compact and extended. For the clear models, the temperature structures are similar at pressure greater than approximately 0.1 bar, but at higher regions in the atmosphere the picket fence GCMs show colder night sides and hotter day sides. Furthermore, the picket fence models no longer have the isothermal upper atmospheres characteristic of the double gray models. Instead, the upper atmospheres show profiles that decrease in temperature with decreasing pressure for the night sides. The dayside of HD 209458b shows a strong temperature inversion, since the picket fence radiative transfer can capture the upper atmosphere heating due to absorption of starlight by TiO and VO. For the cloudy models, the two different radiative transfer schemes produce similar night side temperatures, but the picket fence scheme produces hotter day sides, due both to the heating from radiative transfer and the nature of feedback with the clouds. \textcolor{black}{Note that some of the atmospheric profiles show significant abrupt changes in temperature with pressure. Work by \citet{Beltz2022} using the RM-GCM demonstrated that the implementation of sponge layers at the top of the atmosphere (see that work for description and references) could smooth the profiles, but the models we present here have not included sponge layers, so we see the influence of some numerical noise.}

Optically thick radiatively active clouds have a significant impact on the atmospheric structure of our models, particularly in the upper atmosphere, at pressures less than approximately 0.1 bar. In both HD 189733 b and HD 209458 b, clouds lead to large thermal inversions in the double gray models, primarily along the western terminator where cooler gas advected from the nightside allows cloud formation; these clouds absorb starlight and prevent it from heating deeper regions, causing inverted temperature profiles \citep[as elaborated on in][]{Harada2021}. In the picket fence models, clouds lead to thermal inversions on the cooler and cloudier HD 189733 b models, but have a less pronounced effect on HD 209458 b, which already displays large thermal inversions in the clear case and has less cloud coverage due to the higher temperatures. \textcolor{black}{On the nightside and at pressures less than approximately 10$^{-2}$ bar, clouds decrease non-equatorial temperature by several hundred Kelvin, as seen in Figures \ref{fig:PT-HD189} and \ref{fig:PT-HD209}.}

Figure \ref{fig:PT-HD209} shows that the picket fence models of HD 209458 b have thermal inversions for both the clear and cloudy parameterizations, while the clear double gray models have isothermal upper atmospheres. In the clear picket fence models, there are dayside thermal inversions of hundreds of degrees Kelvin. Furthermore, the thermal inversions in the cloudy picket fence models are several hundred degrees larger than those in the double gray models. The larger day-night differences and presence of inversions are both in agreement with the double gray and picket fence comparisons in \cite{Lee2021}. The more sophisticated picket fence scheme results in larger upper atmosphere opacities and more of the stellar flux being deposited higher in the atmosphere compared to the double gray models. This leads to dayside temperature inversions when the atmosphere is hot enough for TiO and VO to be present and efficiently absorb starlight high in the atmosphere. While HD 189733 b does have some profiles where the temperature increases with decreasing pressure, these effects are caused by dynamics, rather than heating from starlight being absorbed in the upper atmosphere.

Our picket fence radiative transfer approach incorporates contributions from both TiO and VO. However, if these species have condensed into clouds, they should not be present in the upper atmosphere, where they would otherwise generate temperature inversions \citep{Parmentier2013}. Nonetheless, other species may still contribute to such inversions, such as AlO, CaO, NaH and MgH \citep{Gandhi2019}. Although it may not be entirely consistent to include VO clouds while simultaneously assuming that VO exists in the gas phase through the picket fence coefficients, this approach enables us to capture possible thermal inversions. Thermal inversions have not been observed in HD 189733 b \citep{Charbonneau2008, Huitson2012, Crouzet2014}, and although initial broadband photometry suggested thermal inversions on HD 209458 b \citep{Burrows2007, Knutson2008}, more recent high resolution thermal emission observations have disputed thermal inversions in the atmosphere \citep{Diamond2014, Schwarz2015, Line2016}. However, in this work we do not simulate models of HD 209458 b without TiO and VO, as this comparison is outside the scope of this paper, and the primary goal of this work is to understand the impact of different radiative transfer schemes rather than match observations.

While modeling all 13 cloud types intensifies the day-night side temperature differences and thermal inversions, the qualitative effect of radiatively active clouds on the pressure temperature profiles does not change compared to the standard set of cloud species. The largest difference when all 13 species are simulated comes from the addition of absorptive Fe clouds, increasing atmospheric temperatures, especially on the dayside. \textcolor{black}{However, because the 13 cloud models are physically disfavored due to low nucleation rates \citep{Gao2020}, we do not extensively characterize that set of results in this work.} For more details about the differences between the 8 and 13 cloud models see \cite{Roman2021}. In this work, we focus on the 8 species models, the polychromatic nature of clouds, and the differences arising from different cloud vertical extents.

Clouds in the atmospheres of hot Jupiters can thermostat the atmospheric pressure-temperature profile to the species condensation curve temperature, particularly when using double gray radiative transfer, as shown most clearly in Figure \ref{fig:PT-HD209}. \textcolor{black}{This effect in the RM-GCM was first discussed in \citet{Roman2019} regarding Al$_2$O$_3$ clouds thermostating the profiles near the substellar point to the condensation curve temperatures of Al$_2$O$_3$.} Clouds block thermal emission from leaving the planet and raise the local temperature. If the gas heats above the condensation curve temperature the cloud dissipates. The clear atmosphere then allows for more efficient transport of heat out of the planet, and the layer subsequently cools, allowing clouds to re-condense. This feedback loop thermostats the layer temperature near the condensation curve temperature.

\begin{figure*}[!htb]
\begin{center}
\includegraphics[width=0.95\linewidth]{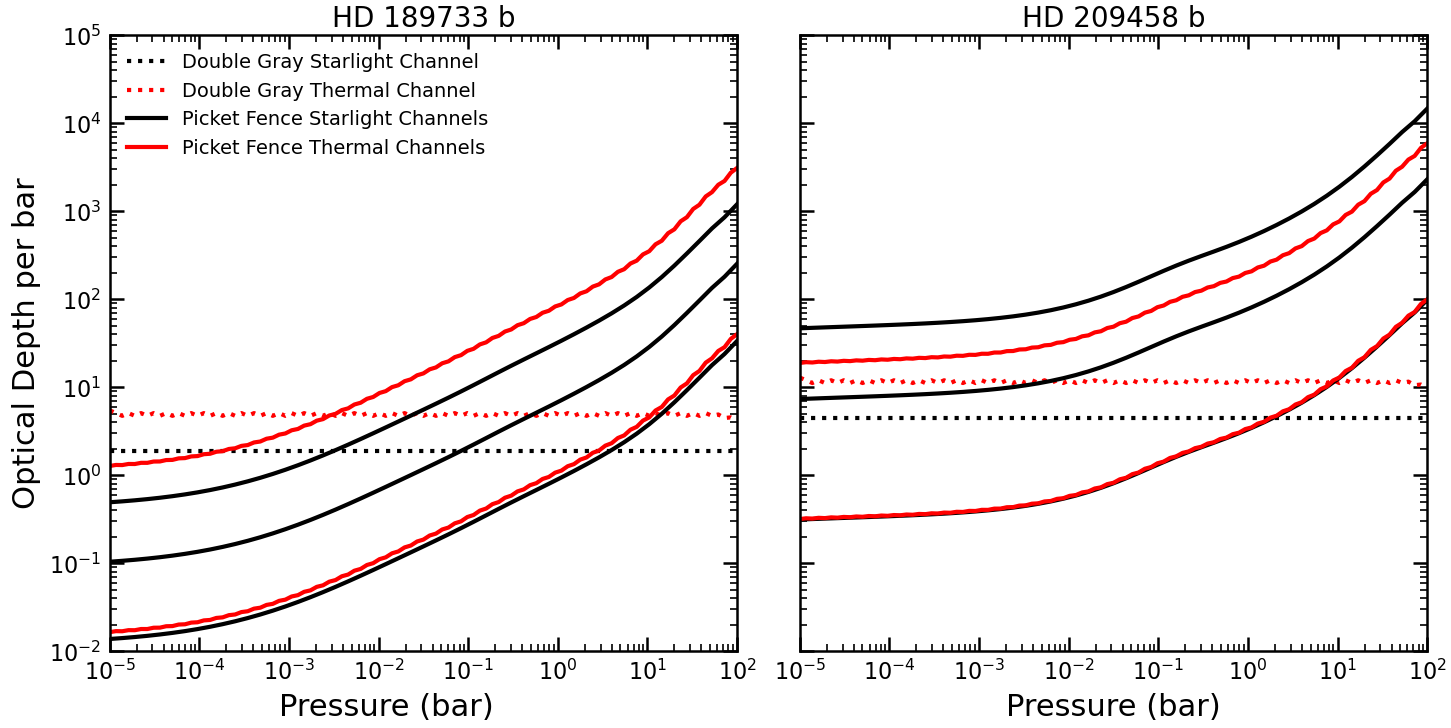}
\caption{\textcolor{black}{The optical depths per bar for HD 189733 b (left) and HD 209458 b (right), evaluated with the analytic \cite{Guillot2010} temperature profiles used for initialization. The solid lines show the picket fence optical depths per bar, and the dotted lines show the double gray optical depths of each layer divided by the layer pressure. Because of constant absorption coefficients, the double gray channels maintain constant optical depths per bar throughout each atmosphere. Depending on the channel number, the picket fence radiative transfer scheme results in optically thinner $and$ optically thicker channels.}}
\label{fig:optical_depths}
\end{center}
\end{figure*}

This \textcolor{black}{thermostating} mechanism is less prominent in the picket fence scheme, where there are multiple radiative transfer channels and the wavelength dependent radiative properties of the clouds change between channels. In the double gray treatment, when clouds form or evaporate, there is an immediate change to the radiative transfer through that region and this can result in strong heating or cooling; with picket fence, this change can be mediated as the gas opacities also adjust with the changing temperature. \textcolor{black}{This thermostating can be seen in the double gray extended cloud models in Figure \ref{fig:PT-HD189}, and the double gray compact and extended cloud models in Figure \ref{fig:Averaged-Clouds-HD209}. However, in comparable picket fence models, especially for HD 209458 b, the temperatures exceed the cloud condensation curve temperatures and the Al$_2$O$_3$ clouds are burnt off. Figure \ref{fig:optical_depths} shows differences in optical depths at each band for the picket fence and double gray clear models of HD 189733 b and HD 209458 b. In the hotter HD 209458 b, TiO and VO increase the gas opacities, leading to more incident starlight deposited at pressures less than 10$^{-3}$ bar, and creates hotter planet daysides. These higher gas opacities allow resulting in absorption and subsequent heating that drowns out the relatively weaker effect of the aerosol heating, reducing the sensitivity to aerosols. This leads to a `break-through' effect for the picket fence HD 209458 b models shown in Figure \ref{fig:PT-HD209}, and only in these models do we find temperatures above the hottest condensation curve temperatures (apart from numerical effects).}

\begin{figure*}[!htb]
\begin{center}
\includegraphics[width=0.95\linewidth]{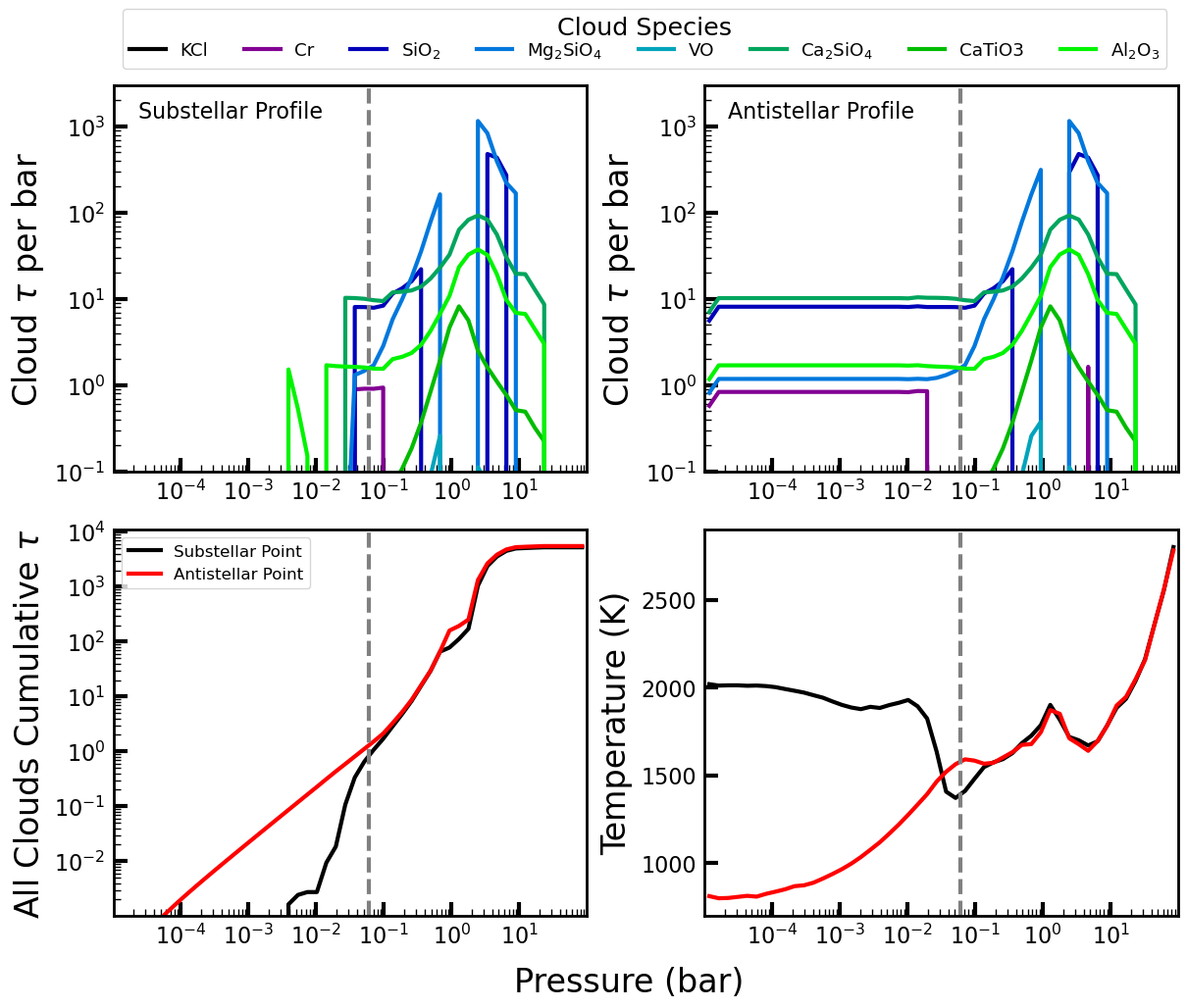}
\caption{Profiles of HD 209458 b with picket fence radiative transfer and extended standard clouds. From top left and clockwise the panels correspond to the cumulative cloud optical depths at $\sim$ 5.0 $\mu m$ of the substellar profile, the anti-stellar profile, the corresponding temperature pressure profiles of both profiles, and the cumulative optical depths of both profiles. The vertical gray dotted line in all panels show the pressure level of the IR photosphere for the clear double gray version of this planet. \textcolor{black}{For the top two panels, the cloud $\tau$ per bar is the optical depth of the condensate layer, divided by the pressure of the atmospheric layer}. The profiles show that radiatively active clouds contribute significantly to the optical depth of the upper atmosphere. }
\label{fig:Aerosol_Profiles_HD209-Table1-Ya-Clouds-Thic-Nuc}
\end{center}
\end{figure*}

Figure \ref{fig:Aerosol_Profiles_HD209-Table1-Ya-Clouds-Thic-Nuc} shows the dayside cumulative cloud optical depths for the picket fence standard extended cloud version of HD 209458 b for two different equatorial profiles: one at the substellar point and one at the anti-stellar point. We chose to show HD 209458 b as it has larger day-night temperature differences (resulting in more inhomogeneous cloud formation) and the extended model because the cloud effects are more pronounced. The colder anti-stellar profile shows more pervasive optically thick clouds. This is particularly important near the photosphere, because upper atmosphere clouds can manifest in the observable emission spectra. Furthermore, these profiles show the relative importance of each cloud species through its cumulative optical depth, which is a function of how much cloud material is present and its radiative properties.

\subsection{Isobaric Structure}
Isobaric temperature and wind maps from our models show the features of hot Jupiter upper atmospheres standard across GCMs. Namely: all of our simulations show day-night side temperature differences of several hundred Kelvin, a strong eastward equatorial jet that weakens past the mid-latitudes, and a hot-spot that has been advected east of the substellar point. While these qualitative features are standard across all of our models, significant differences arise based on which parameterization (double gray vs. picket fence, or cloud distribution) we choose. Deeper into the atmosphere (pressures greater than 10 bar), the differences between the picket fence and double gray schemes are less pronounced.

\begin{figure*}[!htb]
\begin{center}
\includegraphics[width=0.95\linewidth]{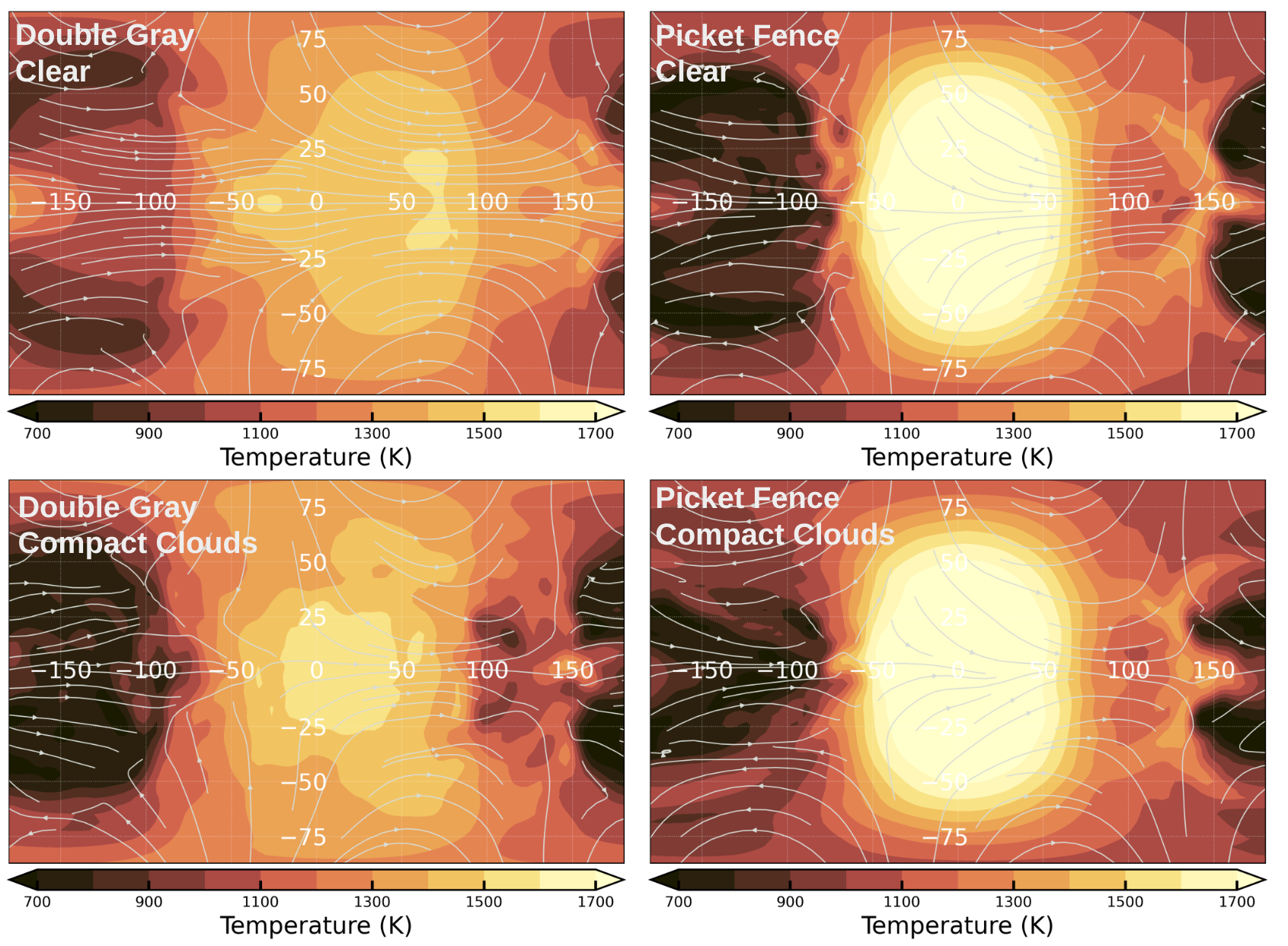}
\caption{The temperature and wind structure of HD 209458 b at a pressure level of 1 mbar. The picket fence radiative transfer results in hotter daysides, especially high in the atmosphere. In addition, whereas in the double-gray models a thermostat effect keeps the dayside temperature profiles near condensation curves and so clouds can form over much of the dayside, the picket fence cloudy model avoids this effect and the dayside temperature profiles exceed the condensation curve, keeping much of the dayside clear.}
\label{fig:Isobars-1mbar}
\end{center}
\end{figure*}

The largest differences between the picket fence and double gray models occur at pressures $\lesssim$ 0.1 bar, due to differences in optical depth structure predicted by each model. Most strikingly, the picket fence models have day sides that are up to 500 K hotter than the double gray models. Figure \ref{fig:Isobars-1mbar} shows the isobaric projection of HD 209458 b at 1 mbar for the double gray and picket fence models with and without compact clouds. 
As discussed above, the picket fence radiative transfer results in hotter daysides and prevents the thermostating effect seen in the double gray models, instead allowing the substellar profiles to reach temperatures above the hottest condensation curve. This keeps the substellar region of the planet cloud-free, resulting in a more similar dayside temperature structure between the clear and cloudy models than is seen for the double gray versions, where the dayside remains largely cloudy, even high in the atmosphere. \textcolor{black}{The clear picket fence models also show disruption in the wind structure near the terminator zones, producing a less coherent equatorial jet with greater longitudinal variation in wind speed compared to the clear double gray models, as shown in the flow streamlines in Figure \ref{fig:Isobars-1mbar}. In the double gray models, a wide coherent jet spans across all longitudes, encircling the planet. However, in the picket fence model, a wide coherent jet is only present across a portion of the dayside. At longitudes of approximately \( -50^{\circ} \) and \( 130^{\circ} \), there is significant north-south flow accompanied by corresponding changes in the coherence of the east-west flow.}

\begin{figure*}[!htb]
\begin{center}
\includegraphics[width=0.75\linewidth]{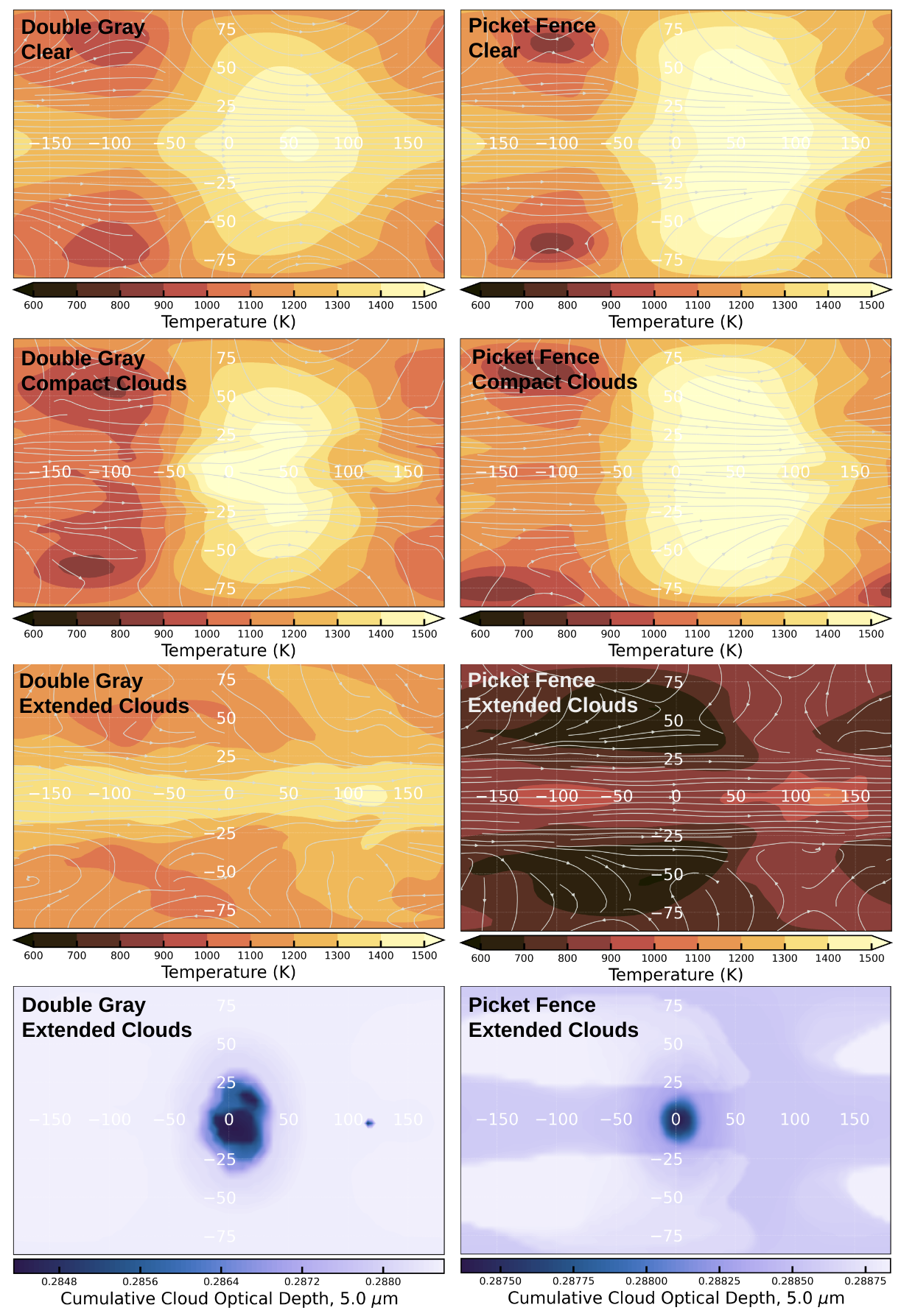}
\caption{The temperature, wind, and cloud structure of HD 189733 b at a pressure level of 141 mbar (top three rows). This is near the infrared photosphere for the double gray, clear model. The final row show plots of cumulative cloud optical depths down to 141 mbar from the top of the atmosphere for the extended cloud models. Although qualitatively similar, the picket fence models differ from the double gray ones in detail and these differences increase when clouds exist in the models.}
\label{fig:Isobars-HD189}
\end{center}
\end{figure*}

\begin{figure*}[!htb]
\begin{center}
\includegraphics[width=0.75\linewidth]{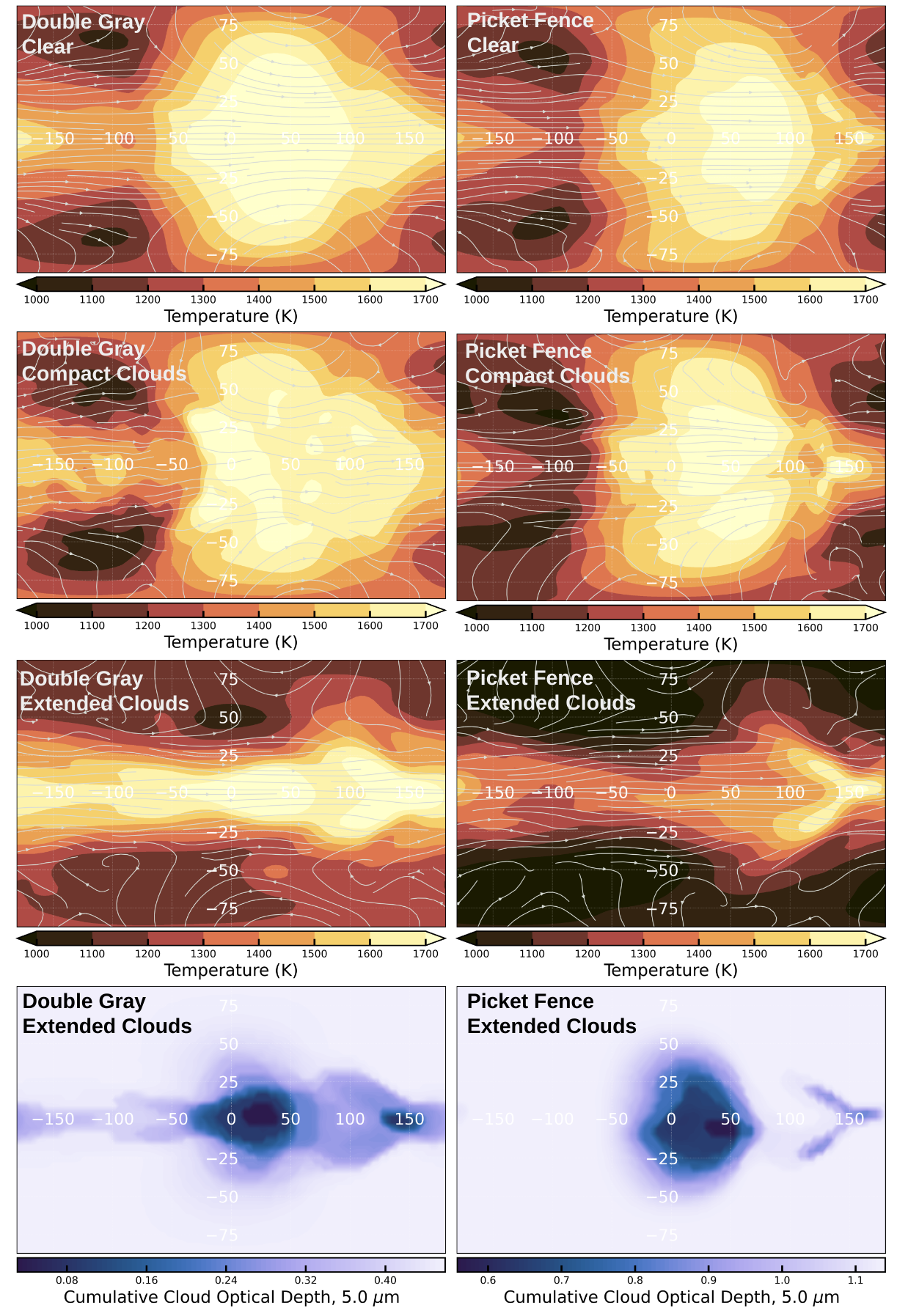}
\caption{The temperature, wind, and cloud structure of HD 209458 b at a pressure level of 60 mbar. This is near the infrared photosphere for the double gray, clear model. At this pressure level, the picket fence models show colder peak dayside temperatures. Particularly in the clear case, the picket fence models shows temperatures near the substellar point several hundred degrees colder than the double gray model.}
\label{fig:Isobars-HD209}
\end{center}
\end{figure*}

Characterizing the near-photospheric pressures of these planets is important for interpreting emission spectra because it corresponds to the region where most of the photons that make up the continuum flux originate. However, only the clear double gray models have a single pressure level corresponding to the IR photosphere. The more complex picket fence models have opacities that vary with the local temperature and pressure fields. Figures \ref{fig:Isobars-HD189} and \ref{fig:Isobars-HD209} show the atmospheric structure at 141 mbar and 60 mbar (the IR photosphere for the clear double gray models) for different models of HD 189733 b and HD 209458 b respectively. \textcolor{black}{In cloudy models of HD 209458 b optically thick ($\tau$ $\gtrsim$ 1) clouds block the emergent flux inhomogeneously over the planet surface at 5 microns. For HD 189733 b, despite pervasive (but inhomogenous) cloud coverage the cumulative cloud optical depth is less than 0.3 at 5.0 $\mu$m. This difference of whether the clouds coverage have become optically thick by the photosphere amplifies differences between extended cloud models of HD 189733 b and HD 209458 b.}

In the extended cloud cases (with implied strong vertical mixing) there are clouds over almost the entire isobaric surface near the clear IR photosphere for both planets. In contrast, the compact cloud cases show nearly clear upper atmospheres. Some species, such as Al$_2$O$_3$, condense at such high temperatures that they are present nearly throughout the entire atmosphere in our extended cloud parameterizations. In the case of HD 189733 b at the IR photosphere, the atmosphere is cold enough that a number of cloud species condense homogeneously. Namely: Al$_2$O$_3$, Ca$_2$SiO$_4$, CaTiO$_3$, Mg$_2$SiO$_4$, and SiO$_2$ blanket the entire isobaric surface; Figure \ref{fig:Isobars-HD189} shows homogeneous cloud coverage for both the double gray and picket fence models, with clouds only dissipating near the substellar point. In the case of HD 209458 b, although the picket fence models have hotter upper atmospheres, the double gray models are hotter near the IR photosphere. Figure \ref{fig:Isobars-HD209} shows that while both the double gray model and the picket fence model are hot enough to dissipate clouds near the substellar point, only the double gray models show a decrease in cloud coverage that extends to the nightside, along the equator. Furthermore, the picket fence models have optically thicker clouds due to the more pervasive formation of Al$_2$O$_3$, Ca$_2$SiO$_4$, and CaTiO$_3$ clouds.

\subsection{Cloud Distributions}
\begin{figure*}[!htb]
\begin{center}
\includegraphics[width=0.95\linewidth]{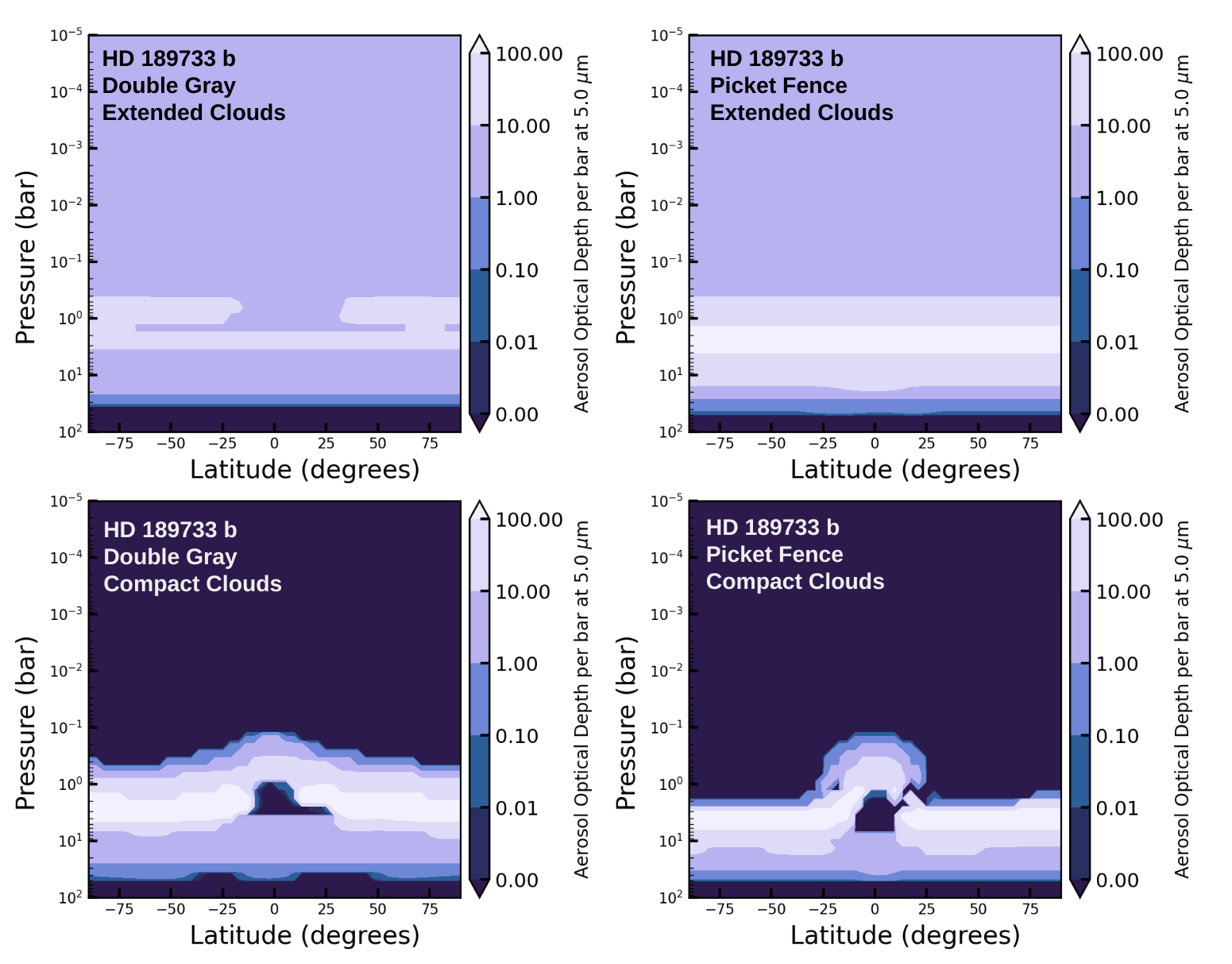}
\caption{The longitudinally averaged cloud structures. The colder HD 189733 b has ubiquitous clouds for the extended case, and deep atmosphere clouds for the compact case.}
\label{fig:Averaged-Clouds-HD189}
\end{center}
\end{figure*}

\begin{figure*}[!htb]
\begin{center}
\includegraphics[width=0.95\linewidth]{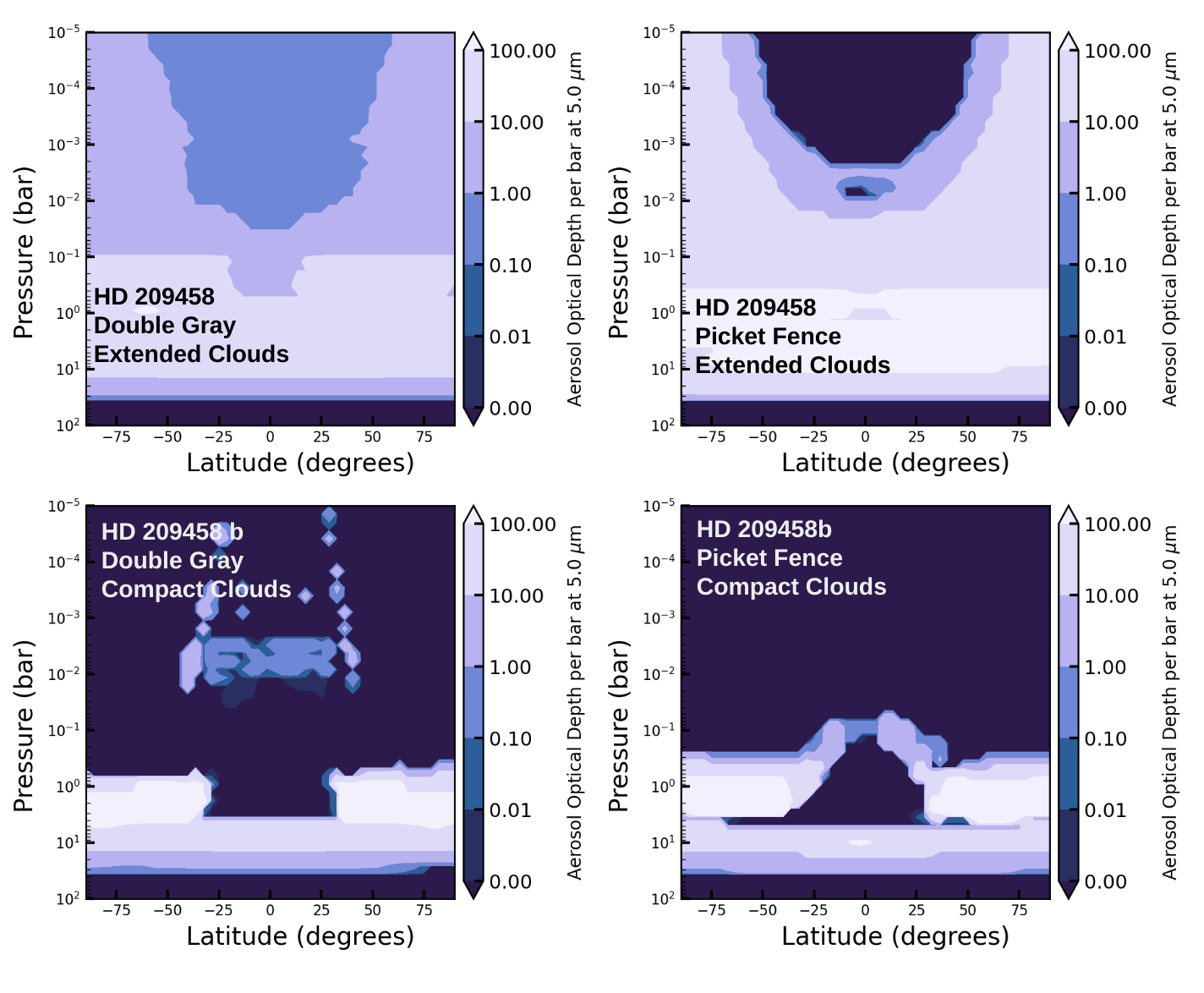}
\caption{The longitudinally averaged cloud structures. In both the picket fence and double gray renditions of HD 209458 b, high-latitude clouds are features in the extended models. However, the picket fence model exhibits temperatures high enough to completely disperse far more of the upper atmosphere equatorial clouds.}
\label{fig:Averaged-Clouds-HD209}
\end{center}
\end{figure*}

Figures \ref{fig:Averaged-Clouds-HD189} and \ref{fig:Averaged-Clouds-HD209} show zonally averaged cloud distributions for a range of cloud parameterizations on HD 189733 b and HD 209458 b. In the deeper atmosphere (from $\sim$30-100 bar) HD 189733 b is hot enough that no clouds form for any cloud parameterizations. For both cloud parameterizations and both radiative transfer schemes, thick clouds formed from approximately 30 bar to 5 bar. Our compact cloud parameterizations show clouds forming deep in the atmosphere but nearly clear upper atmospheres. This is due to the cloud formation being limited to 1.4 scale heights from the bottom most layer for each cloud species. We find that deep in the atmosphere clouds extend over all latitudes for both the double gray and picket fence models, as the day-night differences are muted at the pressures where the compact clouds form. These clouds are generally too deep in the atmosphere to directly manifest in emission or transmission spectra.

Another difference in the longitudinally averaged cloud coverage between the planets is the decrease in equatorial clouds for the picket fence extended cloud model of HD 209458 b, as shown in Figure \ref{fig:Averaged-Clouds-HD209}. This difference stems from the fact that more stellar energy is absorbed in the upper atmosphere in the multiwavelength picket fence method, resulting in temperatures above 2000 K, as shown in Figure \ref{fig:PT-HD209}. Although clouds form in the upper atmospheres near the poles for both the double gray and the picket fence versions of this planet, the picket fence model has thinner clouds for latitudes between -50$^\circ$ and 50$^\circ$. The double gray models still have equatorial clouds in the upper atmosphere of HD 209458 b, primarily Ca$_2$SiO$_4$, CaTiO$_3$ and Al$_2$O$_3$. The dissipation of equatorial cloud is particularly important when interpreting reflected light phase curves (discussed below), as contributions from clouds at these locations can dominate the reflected flux due to geometric effects.

The larger spatial differences in cloud coverage between the double gray and picket fence models are also important for interpreting observations. Especially for planets at different viewing inclinations, spatial inhomogeneities in temperature and cloud coverage will manifest in emission spectra, transmission spectra, and phase curves \citep{line2016influence,powell2019transit,harada2021signatures,Malsky2021,feinstein2023early,savel2023limb}. \textcolor{black}{Recent work has show that exoplanets with large day-night temperature contrasts and other spatial inhomogeneities can result in systematic biases in retrievals \citep{Caldas2019, Taylor2020, Pluriel2020, MacDonald2020}. The large spatial inhomogeneities found in these models, particularly for the cloudy simulations, show the importance of 2D and 3D retrieval frameworks.}

\subsection{Zonal Wind Structures}
\begin{figure*}[!htb]
\begin{center}
\includegraphics[width=0.9\linewidth]{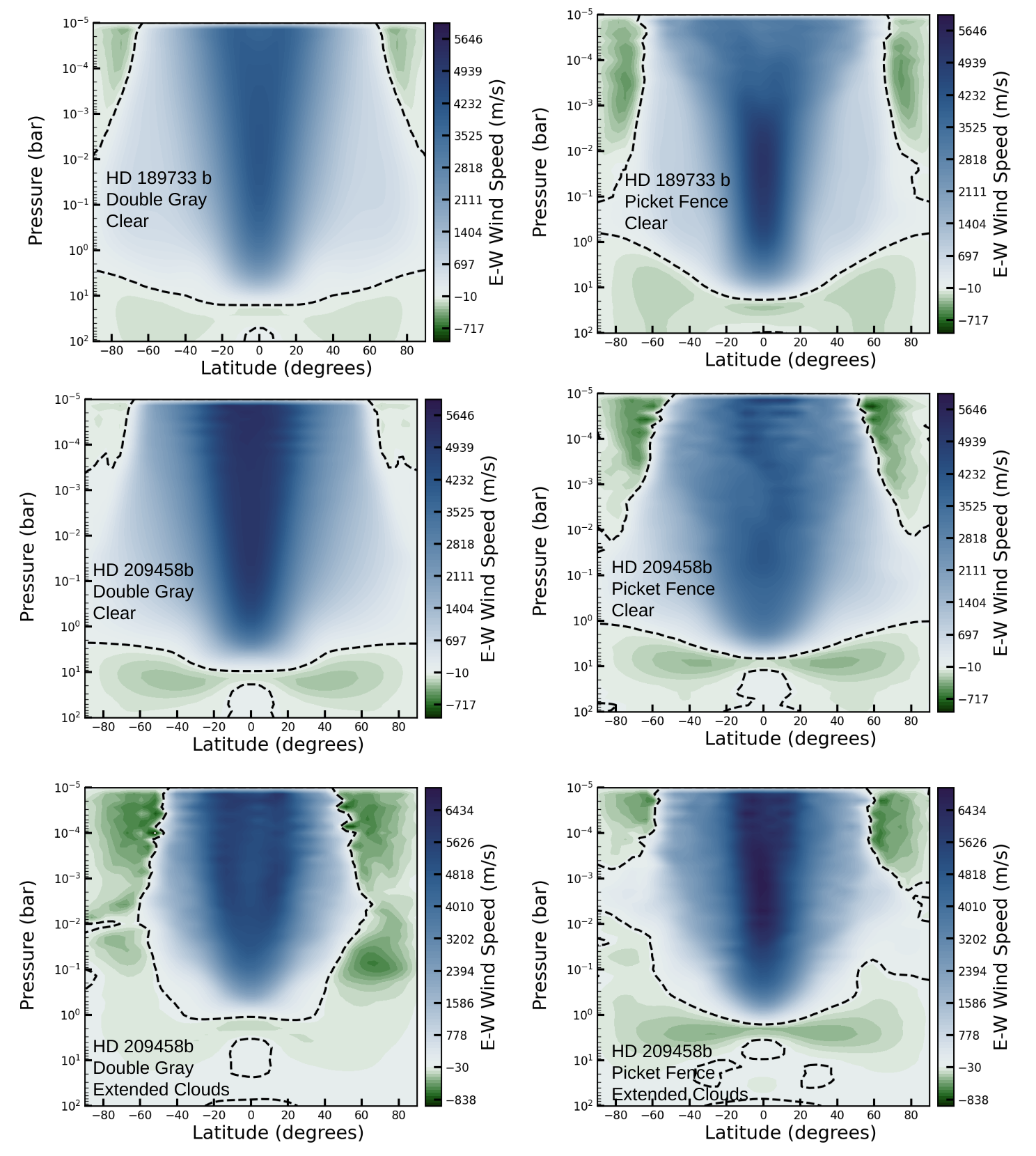}
\caption{The longitudinally averaged zonal wind speeds of HD 189733 b without clouds (top row), HD 209458 b without clouds clouds (middle row), and HD 209458 b with extended clouds (bottom row). The models were run using double gray radiative transfer (left column) and the new picket fence scheme (right column). The dark blue central feature corresponds to the eastward equatorial jet. The black dotted line corresponds to a contour where the zonally averaged wind speed is 0 km s$^{-1}$. All models show strong eastward equatorial jets, standard in hot Jupiter circulation patterns, but the strength of the jet and its spatial extent differ between the double gray and picket fence models, especially when there are clouds present.}
\label{fig:Winds}
\end{center}
\end{figure*}

Qualitatively, we find similar atmospheric wind structures for both the double gray and picket fence GCMs. Figure \ref{fig:Winds} shows the longitudinally averaged zonal wind speeds of HD 209458 b and HD 189733 b. All of our models showed common features for the mid to deep atmosphere winds: an eastward equatorial jet with speeds in excess of 4 km s$^{-1}$ and extending down to at least 1 bar, and significantly slower mid-latitude winds.

Although the polychromatic scheme results in larger upper atmosphere day-night temperature differences than the double gray version, it also leads to winds that disrupt and slow the equatorial jet near the terminator \textcolor{black}{(as discussed above and shown in Figure \ref{fig:Isobars-1mbar})}, particularly in the case of HD 209458 b. Figure \ref{fig:Winds} shows the longitudinally averaged zonal wind speeds of a set of our models. The clear picket fence model for HD 189733 b has peak wind speeds about 1 km/s faster than the double gray model at pressures larger than $\sim$1 bar. However, the zonally averaged wind speeds show a faster equatorial jet for the double gray model of HD 209458 b, despite smaller temperature differences than the picket fence model. This is because the zonal averaging is obscuring a more complex wind pattern. The picket fence and double gray models both have peak wind speeds near the equator of approximately 6 km/s. The introduction of extended clouds increases day-night temperature differences, and increases the top wind speeds in both schemes. Furthermore, the addition of clouds adds more small scale features to the atmospheric temperature and circulation patterns, especially in the picket fence models. All models show slow westward winds in the deep atmosphere and near the poles at pressures lower than 0.01 bar.

\subsection{Phase Curves}
\begin{figure*}[!htb]
\begin{center}
\includegraphics[width=0.9\linewidth]{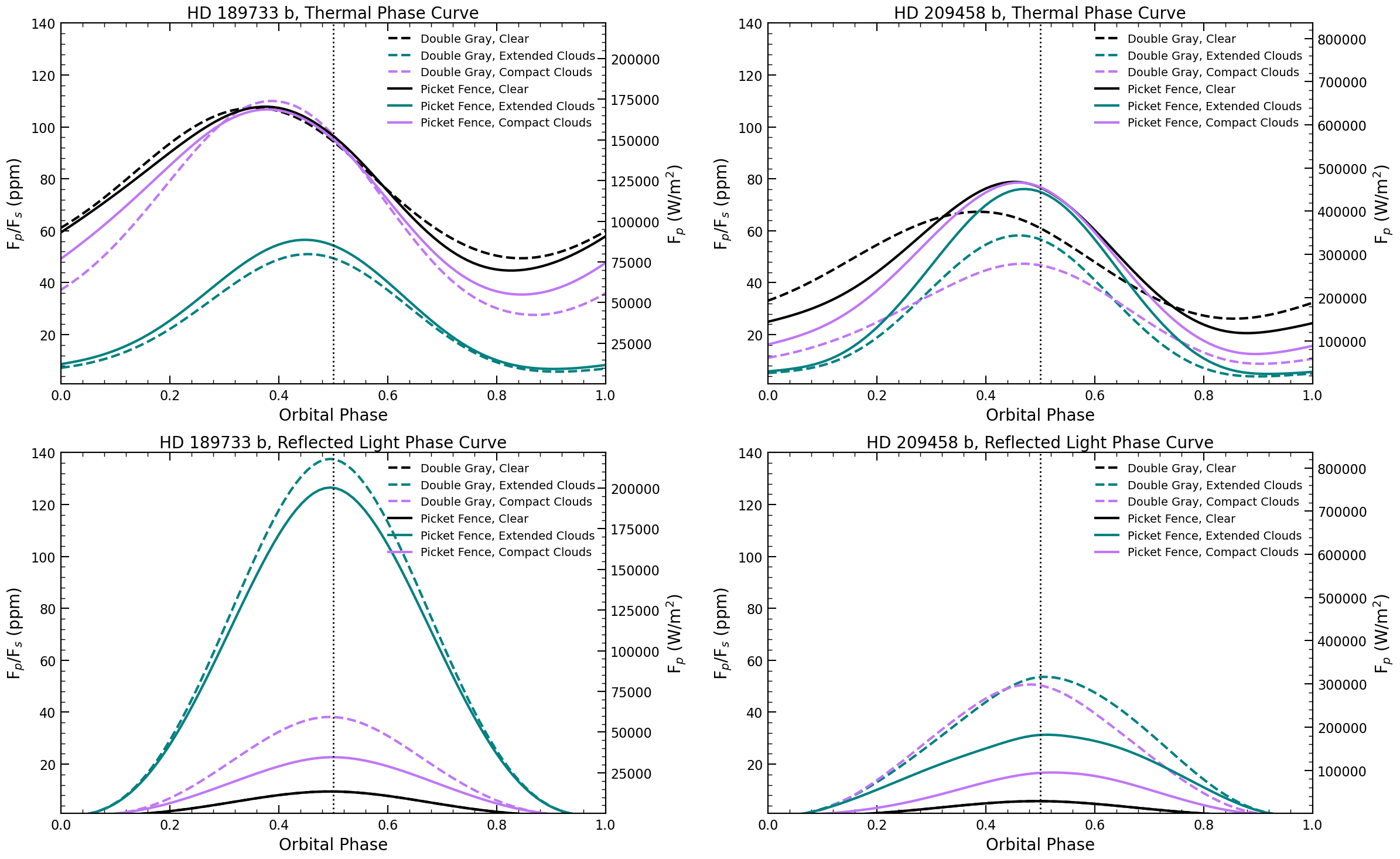}
\caption{Top, the thermally emitted broadband phase curves. Bottom, the reflected broadband phase curves. These are the fluxes directly from the GCM radiative transfer, capturing the entire thermal and starlight bands. \textcolor{black}{Dotted lines show the double gray radiative transfer, solid lines show the picket fence results. The colors show different cloud parameterizations: clear (black), compact clouds (purple), and extended clouds (teal). The choice of a radiative transfer scheme and the vertical extent of clouds can be distinguished based on phase curve observations. Using the stellar effective temperatures for HD 189733 b and HD 209458b from Table \ref{tab:planet_parameters}, we compute the bolometric fluxes.}}
\label{fig:phasecurves}
\end{center}
\end{figure*}

The thermal and reflected light phase curves for HD 189733 b and HD 209458 b are shown in Figure \ref{fig:phasecurves}, and Table \ref{tab:bonds} lists the phases of peak flux for each model. These phase curves were created by taking the outward flux from the upper boundary of the GCM at each location and integrating over the visible hemisphere (weighted appropriately by the solid angle) at each orbital phase. Due to the way the radiative transfer is parameterized within the GCM, these are the contributions from the thermal and starlight bands.

Clouds significantly change the overall energy balance in both the double gray and picket fence schemes. Although the clear version of each planet had a prescribed base global Bond albedo of 0.10 from Rayleigh scattering, adding clouds increases the global Bond albedo of the planet, from values of 0.11 to as high as 0.71. The Bond albedos of all models presented here are shown in Table \ref{tab:bonds}. The models with extended clouds are the most reflective and the double gray versions of each cloudy model are more reflective than the picket fence versions, as we will explain in more detail below. We are able to see this increased reflection directly in the reflected light phase curves and indirectly in a decrease in overall flux of the thermal emission phase curves. Furthermore, by comparing the total incident stellar flux and internal flux to the thermal flux and reflected starlight, we were able to benchmark the overall energy conservation within each model. We find outgoing to incoming energy balances were within 3\% for all models, and less than 1\% for the majority of models. These values were only calculated from the final output of each model and so the energy balances may actually be even better when averaged over time.

Similarly to \cite{Parmentier2021} and \cite{Roman2021}, we find that clouds lead to smaller thermal phase curve offsets and larger thermal phase curve amplitudes for both HD 189733 b and HD 209458 b. Clouds affect phase curves through a number of processes. The addition of clouds increases global Bond albedos, decreasing the global energy budget. Conversely, clouds strengthen the greenhouse effect, which in turn warms the atmosphere beneath them. Furthermore, clouds introduce an additional source of opacity and move the photospheric pressure level to higher altitudes, and alter the temperature structure through scattering and absorption. The blanketing can be seen through Figure \ref{fig:Aerosol_Profiles_HD209-Table1-Ya-Clouds-Thic-Nuc}, where clouds have formed at both the substellar point and the antistellar point by the IR photosphere. Locations above the photosphere are generally colder than (or the same temperature as) the photosphere for HD 189733 b, but may be a higher temperature for HD 209458 b models where thermal inversions are pervasive. The exact nature of how clouds manifest in the broadband phase curves is due to the complex combination of the disk integrated emission, as well as the radiative properties of the clouds, the type of clouds forming, and the global energy balances of the planets.

\textcolor{black}{The hotter dayside upper atmospheres and cooler night sides, result in the smaller thermal phase curve offset and larger thermal phase curve amplitude of the clear picket fence model shown in Figure \ref{fig:phasecurves}. Importantly, the picket fence scheme has an extra thermal channel compared to double gray, resulting in more efficient cooling and night-sides hundreds of degrees Kelvin cooler than the double gray scheme.} All models show an eastward thermal peak offset, with the double gray HD 189733 b having the largest offset (peaking at a phase of 0.36) and the picket fence extended cloud HD 209458 b model having the smallest offset (peaking at a phase of 0.48). All model offsets are detailed in Table \ref{tab:bonds}). Generally, the thermal phase curves of the double gray and picket fence models are similar both in terms of amplitude and phase curve offset for HD 189733 b, which is in line with their similar pressure-temperature profiles.

When no clouds are present, the double gray and picket fence reflected light curves match exactly, as the only source of scattering is our imposed Rayleigh scattering. Radiatively active clouds scatter and reflect incident starlight, increasing planetary albedos. The double gray models are more reflective, because they have cooler day side upper atmospheres, resulting in greater cloud coverage and more reflected starlight. This effect is more pronounced for HD 209458 b where the temperature differences between the double gray and picket fence versions are greater. The HD 189733 b models almost all have reflected light phase curves that peak at phases of 0.5. These models are cold enough that clouds form across both the day and the night sides, resulting in mostly homogeneous cloud coverage and reflected light maps. In contrast, the radiative transfer choice had a large effect on the reflected light phase curves of the hotter HD 209458 b. Here, the double gray models had significantly larger reflected light phase curve amplitudes. This is because the picket fence models had day sides hot enough to dissipate all clouds in the upper atmosphere. This resulted in less overall reflected starlight. Furthermore, both the double gray and picket fence models had reflected light phases close to, but not exactly at, 0.5 (0.49 - 0.53), indicative of the inhomogeneous dayside cloud coverage on this planet.

One feature to notice is that the default extended cloud models are more reflective than the extended all-cloud models. This is because the default cloud models exclude highly absorptive Fe clouds, and allow for reflective silicate clouds to form, dominating the optical radiative transfer. This behavior was noted in \cite{Roman2021} for the double gray models, and remains true for the picket fence radiative transfer. As expected, the highly reflective default cloud models also have the smallest integrated thermal emission---a necessary feature to conserve the global energy balance. Similarly, the extended cloud models are more reflective than the compact cloud models, which are in turn more reflective than the clear models. This behavior is matched in the thermal phase curves, where the integrated thermal emission is largest for the clear models, then the compact cloud models, then the extended cloud models. However, the converse is true for the compact cloud models---the default cloud models are less reflective than the extended cloud models, particularly for HD 189733 b. This is due to nuances in the location where different cloud species form and which species form close to the photosphere.

\begin{table*}[!ht]
\centering
\begin{tabular}{ccccc}
\toprule
\multicolumn{5}{c}{HD 189733 b} \\
\midrule
Model & Bond albedo & Reflected phase curve max & Thermal phase curve max & Thermal amplitude \\
\midrule
Double Gray, Clear & 0.10 & 0.50 & 0.36 & 0.54 \\
Picket Fence, Clear & 0.10 & 0.50 & 0.38 & 0.59 \\
Double Gray, Default Clouds, Comp. & 0.17 & 0.50 & 0.39 & 0.75 \\
Picket Fence, Default Clouds, Comp. & 0.11 & 0.50 & 0.38 & 0.67 \\
Double Gray, Default Clouds, Ext. & 0.71 & 0.50 & 0.46 & 0.89 \\
Picket Fence, Default Clouds, Ext. & 0.66 & 0.50 & 0.45 & 0.88 \\
Double Gray, All Clouds, Comp. & 0.32 & 0.50 & 0.41 & 0.82 \\
Picket Fence, All Clouds, Comp. & 0.19 & 0.52 & 0.39 & 0.66 \\
Double Gray, All Clouds, Ext. & 0.34 & 0.50 & 0.49 & 0.95 \\
Picket Fence, All Clouds, Ext. & 0.33 & 0.50 & 0.47 & 0.90 \\
\midrule
\multicolumn{5}{c}{HD 209458b} \\
\midrule
Model & Bond albedo & Reflected phase curve max & Thermal phase curve max & Thermal amplitude \\
\midrule
Double Gray, Clear & 0.10 & 0.50 & 0.39 & 0.61 \\
Picket Fence, Clear & 0.10 & 0.50 & 0.46 & 0.74 \\
Double Gray, Default Clouds, Comp. & 0.40 & 0.49 & 0.47 & 0.81 \\
Picket Fence, Default Clouds, Comp. & 0.14 & 0.53 & 0.47 & 0.84 \\
Double Gray, Default Clouds, Ext. & 0.50 & 0.51 & 0.47 & 0.93 \\
Picket Fence, Default Clouds, Ext. & 0.33 & 0.52 & 0.48 & 0.94 \\
Double Gray, All Clouds, Comp. & 0.41 & 0.50 & 0.46 & 0.84 \\
Picket Fence, All Clouds, Comp.  & 0.16 & 0.54 & 0.46 & 0.85 \\
Double Gray, All Clouds, Ext. & 0.35 & 0.50 & 0.48 & 0.93 \\
Picket Fence, All Clouds, Ext. & 0.18 & 0.51 & 0.47 & 0.89 \\
\bottomrule
\end{tabular}
\caption{The Bond albedos, orbital phases at which the reflected and thermal light from the planet peaks, for all models.}\label{tab:bonds}
\end{table*}

\section{Discussion}\label{sec:Discussion}
In this work, our new contribution is the coupling of a picket fence radiative transfer scheme with radiatively active clouds. This builds upon previous works that have studied various cloud modeling assumptions within GCMs, as well as comparisons between radiative transfer routines within clear models. In the following sections we compare our results to previous literature.

\subsection{Radiative Transfer Comparisons}
In order to benchmark the accuracy of the picket fence model, we compare our results against previous picket fence implementations, as well as against \textcolor{black}{k-distribution} radiative transfer schemes. In particular, we focus on \cite{Lee2021}, who compared double gray, picket fence, and \textcolor{black}{k-distribution} schemes for cloud-free models of HD 209458 b. This model uses a finite volume dynamical core, but slightly different formulations of the primitive equations \citep{Lee2021}. As noted in \cite{Lee2021}, small differences in GCM parameterization can have large effects in the final model simulations, and tracing back the exact cause of model discrepancies is difficult. However, we note the broad agreement between our picket fence models, as well as the common differences we and \cite{Lee2021} found between double gray and picket fence models.

We find similar upper atmospheres as \cite{Lee2021} for HD 209458 b, ranging from approximately 750 K on the nightside to 2000 K on the dayside. Furthermore, we find similar deep atmosphere temperatures as in \cite{Lee2021}. At 10 bar, our models have a temperature of approximately 2000 K, regardless of longitude, matching that of \cite{Lee2021}. We also find monotonically increasing temperatures at depths greater than $\sim$ 10 bar, similar to \cite{Lee2021}. Our isobaric temperature and wind projections are similar to those presented in \cite{Lee2021}. For example, at a pressure level of 0.1 bar we both find global hot-spots that extend past longitudes of 90$^\circ$, an eastward equatorial jet extending from +30$^\circ$ to -30$^\circ$ latitude, and temperatures ranging from 1200 K to 1800 K. One difference, however, we find that the minimum temperatures in our models are at higher latitudes than those shown in \cite{Lee2021}.

We find similar wind structures between our double gray models of HD 209458 b and those presented in \cite{Lee2021}. Both models show super-rotating equatorial jets extending from latitudes of approximately -40$^\circ$ to +40$^\circ$ and slow westward polar and deep atmosphere winds. The model shown in \cite{Lee2021} reaches peak wind speeds of approximately 5 km/s, while the model shown here has peak wind speeds of $\sim$5.5 km/s. Furthermore, the double gray HD 209458 b model presented in \cite{Rauscher2012}, also compared to in \cite{Lee2021}, shows peak wind speeds in excess of 7 km/s, making our results an intermediary between these two models.

Our clear picket fence models show slower peak wind speeds and a different super-rotating equatorial jet structure than the model presented in \cite{Lee2021}. Our models show peak zonally averaged eastward wind speeds of approximately 4 km/s, while the models in \cite{Lee2021} have peak speeds of approximately 5 km/s. Furthermore, our models show a broad equatorial jet from latitudes of approximately -50$^\circ$ to +50$^\circ$, but without a central core of winds in excess of 4 km/s. In juxtaposition, the wind structure of the models in \cite{Lee2021} show a narrower jet, and one with peak wind speeds at $\sim$10$^{-4}$ bar, which then decays deeper into the atmosphere. \cite{Lee2021} also find westward winds at pressures less than 10$^{-5}$ bar. However, our models only extend to this pressure level, so we are unable to compare behavior at lower pressures. Last, both our models and the ones shown in \cite{Lee2021} show many commonalities, such as a slow westward winds at pressures larger than $\sim$ 3 bar and near the poles, and an equatorial jet that extends to $\sim$ 3 bar.

\textcolor{black}{While comparing the overall wind speeds in these models shows some reasonable agreement, we do remind the reader of the discussion in Section \ref{sec:Introduction} about how numerical dissipation can influence wind speeds \citep{Heng2011a}. These models are not using identical numerical dissipation and so that will also factor into their relative winds speeds. In addition, we remind the reader that we may also expect the strength of numerical dissipation appropriate for a model to depend on the radiative timescales in the atmosphere \citep{Thrastarson2011}. So, for example, differences in the radiative transfer implementation in \cite{Rauscher2012} and the \cite{Toon1989} scheme now used in the RM-GCM could therefore introduce nuanced differences in the detailed thermal state of the atmosphere, resulting in complicating consequences for the numerical dissipation and overall wind speeds in these models, even when comparing double-gray versions of each.}

\subsection{Cloud Treatment Comparisons}
We find similar behavior between the cloud treatment presented here and more sophisticated cloud models \citep[e.g.,][]{Lines2018, Parmentier2021, Christie2021, Komacek2022}. Our zonally averaged extended cloud distributions are qualitatively consistent those presented in \cite{Lines2018}, particularly for the hot HD 209458 b model, who modeled HD 189733 b and HD 209458 b with a \textcolor{black}{k-distribution} gas radiative transfer routine and a kinetic cloud formation scheme. We both find extensive cloud coverage from the deep atmosphere to near the top of the atmosphere, with the densest clouds near the poles and at pressures below 10$^{-2}$ bar. Additionally, both our models show a dispersal of clouds deep into the planet atmosphere near the equator, due to the higher temperatures. However, the standard HD 209458 b model in \cite{Lines2018} shows clouds at pressures larger than 30 bar for both  HD 189733 b and HD 209458 b, while our models show no clouds due to the high deep atmosphere temperatures because of our choice of a high T$_{int}$. Last, our isobaric projections of the extended cloudy models of HD 189733 b and HD 209458 b show pervasive cloud coverage across the nightside of the planet, including at equatorial latitudes. In contrast, the models in \cite{Lines2018} show cloud particle number densities close to zero for latitudes between +15$^{\circ}$ and -15$^{\circ}$ for the HD 209458 b models at similar pressures. This behavior is closer to that of our compact cloud parameterizations, but still show significant differences in the extent and region of cloud formation.

Our models closely match the results of \cite{Roman2021}, due to the shared heritage of our models. \cite{Roman2021} ran double gray models with radiatively active clouds over a range of temperatures. The more detailed differences between the results of \cite{Roman2021} and our models are reflected in our comparisons between the double gray and picket fence models presented previously. Further differences between models stem from a different Rayleigh scattering parameterization, as well as updated cloud radiative properties.

In agreement with previous models \citep[e.g.,][]{Roman2019,Roman2021, Parmentier2021}), we find that the presence of clouds on the nightside generally increased peak phase curve amplitudes and decreased phase curve offsets. \cite{Parmentier2021} additionally found that the precise effect of clouds on observables was complex, and were effected by cloud chemical composition and particle size---two parameters with large \textit{a priori} uncertainty. \cite{Christie2021} find that the relatively unconstrained sedimentation factor (or cloud vertical scale) can have a large impact on radiative feedback and atmospheric dynamics. Our models show similarly large differences between models with compact or extended cloud parameterizations. The cloud parameterizations presented here spanned between approximately 1.4 scale and 16 scale heights (the compact and extended cloud parameterizations), and show the need to explore cloud vertical extent. Last, \cite{Komacek2022} explore radiatively active clouds on ultra-hot Jupiters and find nightside clouds that warm the region below the cloud deck through the greenhouse effect. Because our clouds either form throughout the entire atmosphere, or at pressures larger than approximately 1 bar (depending on the compact vs extended parameterization), this blanketing effect is muted. Exploring a wider range of equilibrium temperatures and cloud extents may show more parallels between our results and the findings from \cite{Komacek2022}. \textcolor{black}{Finally, although it is outside the scope of this work, a future study could evaluate whether the vertical extents assumed in this work are consistent with the extents predicted by microphysical models, given the same atmospheric pressure-temperature profiles. Although this approach would neglect the importance of horizontal advection, it would still be informative regarding whether the assumed vertical extents are realistic.}

\section{Conclusions}\label{sec:Conclusions}
In this work we present General Circulation Models with picket fence radiative transfer and radiatively active clouds as a novel method in numerical simulations of hot Jupiters. The picket fence radiative transfer method allows us to recreate the results of far more complex schemes, but at a fraction of the computational cost. When this method is used in concert with cloud condensation we can gain further insight into the complex feedback between clouds and atmospheric structure and the polychromatic nature of hot Jupiter atmospheres.

Overall, the picket fence routine is an efficient and accurate radiative transfer method that better captures the multiwavelength effects of hot Jupiter atmospheres. Compared to double gray, these effects include breaking through the cloud blanketing effect, more efficient thermal cooling of the upper atmospheres, and the larger upper atmosphere starlight opacities modeling the deposition of stellar flux more accurately. As discussed above, double gray models compare favorably for cooler planets, where they qualitatively match the results of the picket fence models. A fully multi-wavelength \textcolor{black}{k-distribution} treatment would be necessary in cases where more complex Rayleigh scattering effects are needed, or where the radiative properties of the atmospheric aerosols vary significantly with wavelength.

We characterize the differences between models that use picket fence and double gray radiative transfer with radiatively active clouds, and find a number of key interactions:

\begin{enumerate}
    \item The picket fence scheme creates larger day-night temperature differences than the double gray scheme in the upper atmosphere of clear models, particularly for the more irradiated HD 209458 b. Furthermore, the temperature profiles in the cloudy double gray models experience a thermostating effect that aligns them with the hottest condensation curve on the dayside \citep[as discussed in ][]{Roman2019}, but the multi-band radiative transfer in the picket fence models diminishes this effect and allows for hotter and clear dayside regions.
    \item While the picket fence approach yields more significant day-night temperature disparities in the upper atmosphere compared to the double gray method, the zonally averaged wind speeds are slowed, due to less coherent flow in the equatorial jet as it passes over the terminators. In both double gray and picket fence configurations, incorporating extensive clouds intensifies day-night temperature differences and increases the peak wind speeds.
    \item 
    Upper atmosphere equatorial and mid-latitude clouds are less pervasive with polychromatic radiative transfer in the case of HD 209458 b, due to larger gas opacities and the resulting larger dayside temperatures. Furthermore, the inhomogenous dissipation of dayside clouds for our HD 209458 picket fence models leads to larger thermal phase curve amplitudes and smaller reflected light curve amplitudes.
    \item Clouds considerably alter the total energy balance in both double gray and picket fence configurations. The introduction of extended clouds increased the planetary global Bond albedos to at least 0.33, and up to 0.71 depending on cloud extent and radiative transfer scheme. Picket fence models resulted in lower Bond albedos than their double gray counterparts because of the larger picket fence upper atmosphere gas opacities. Furthermore, radiatively active clouds decrease phase curve offsets and increase phase curve amplitudes for both double gray and picket fence radiative transfer schemes, as found in previous works.
\end{enumerate}

\textcolor{black}{The exquisite precision measurements available from} JWST \textcolor{black}{and ground-based high-resolution spectroscopy} are revolutionizing atmospheric characterization by probing inherently 3D atmospheric properties \textcolor{black}{\citep[e.g.,][]{Ehrenreich2020,Beltz2021,Herman2022,Pino2022,Prinoth2022,Coulombe2023,Kempton2023,vanSluijs2023}}. In the upcoming decade, new high-resolution spectrographs on 30-m class telescopes will allow for the analysis of spectra from a plethora of exoplanets and a greater understanding of the 3D nature of hot Jupiters. Numerical models are critical to understanding the dynamics of hot Jupiter atmospheres and understanding how the structure of these planets manifests in observables. The diversity of exoplanets and the extensive variety of conditions in which they exist shows the need for efficient, explainable GCMs that can model a range of physical mechanisms in concert.

\acknowledgements
IM would like to thank Marianne Cowherd for providing editorial suggestions on drafts of this manuscript, as well as NC and CC for their support. Sebastian Wolf wrote the Mie theory code used in this work. This research was supported in part through computational resources and services provided by Advanced Research Computing at the University of Michigan, Ann Arbor. This work received financial support from the NASA Exoplanets Research Program Grant \#80NSSC22K0313 and the Heising-Simons Foundation.

\clearpage
\newpage

\appendix\label{appendix}
\section{Rayleigh Scattering}
As part of the improvements detailed in this paper, we also implement a modification to how the model simulates Rayleigh scattering. These results apply to both the cloudy models and the clear models. In the clear models, Rayleigh scattering is the only source of scattering and the sole cause of reflected starlight. For our implementation of Rayleigh scattering within the RM-GCM, the overall Bond albedo can be set to match observations and/or more sophisticated models. However, there exists significant uncertainty about the global Bond albedos for the population of planets we wish to model. For example, \cite{Cahoy2010} calculate Bond albedos of approximately 0.4 for clear Jupiter-like planets at 0.8 AU. In contrast, \cite{Parmentier2015} approximates the Bond albedo for clear solar composition planets and finds values from $\sim$0.02 - 0.2 for planets with temperatures from 1000 - 2000 K.

Previous double-gray versions of the RM-GCM implemented Rayleigh scattering by calculating the atmospheric attenuation due to molecular scattering \citep{Roman2017}. Briefly, this scheme assumed an atmosphere of H$_2$, He, and H$_2$O and computed Rayleigh scattering optical depth per bar for an H$_2$ gas following \cite{Dalgarno1962}. Each molecular species contributes to the total Rayleigh scattering based on its atmospheric mole fraction and intrinsic index of refraction. However, this method performs poorly when combined with our picket fence radiative transfer, as the picket fence is tuned to approximate the overall net transfer of radiation and heating rates. Attempting to pick any wavelength at which to evaluate the Rayleigh scattering (or even any three wavelengths for all the starlight channels) produces unphysical scattering effects (such as an increase in reflected light near the terminator) for some subset of hot Jupiters that we wish to model. Individually tuning the Rayleigh scattering for each planet is also not desirable, as it leads to over-parameterization between models.

In order to implement Rayleigh scattering within the new picket fence framework, we make several changes to the RM-GCM. First, we updated the radiative transfer to keep the functionality of the previous Rayleigh scattering methods if desired. This required evaluating the Rayleigh scattering at three wavelengths (for the three starlight channels), instead of one. If the wavelengths that Rayleigh scattering is evaluated at are identical for each of the starlight channels this method can simplify to a double gray radiative transfer. Next, we update an alternative scattering parameterization (used in this paper) where Rayleigh scattering is not calculated as a optical depth contribution, and instead each profile has an imposed top-of-the-atmosphere Bond albedo that reduces the incident stellar flux. This method, although simple, is able to correctly model the energy balance of the planet while also implementing the picket fence optical depth calculations.

\begin{figure}[!htb]
\begin{center}
\includegraphics[width=0.8\linewidth]{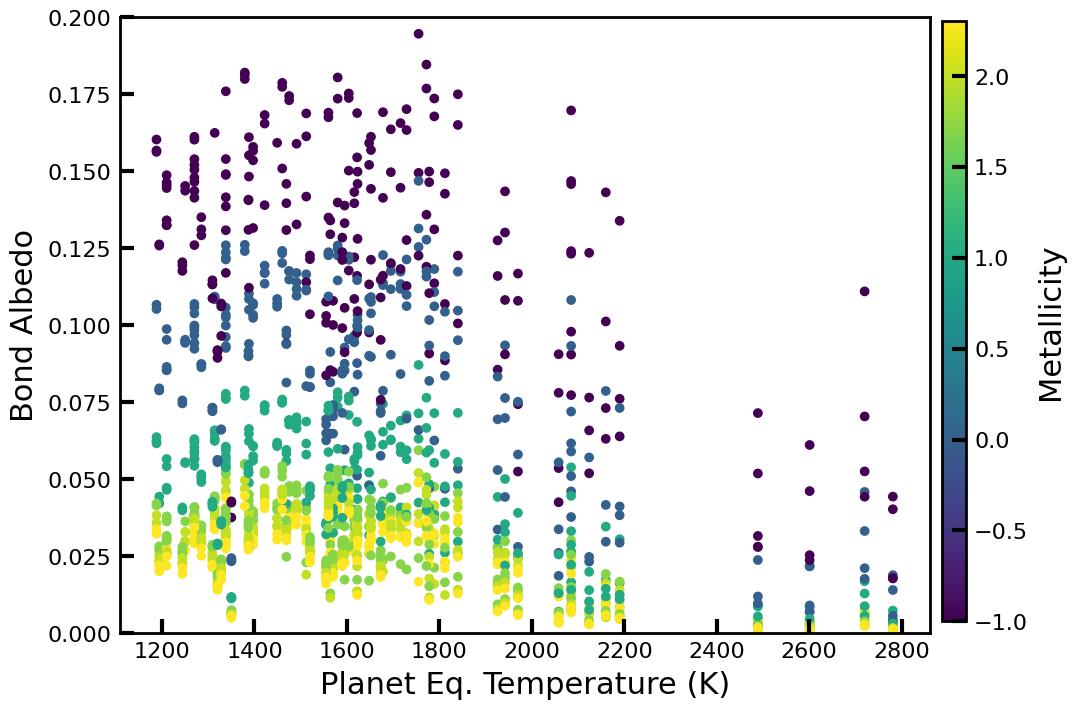}
\caption{The Bond albedos calculated for planets in the grid of models from \cite{Goyal2020}, using the multiwavelength radiative transfer code PICASO.}
\label{fig:bond_albedo_grid}
\end{center}
\end{figure}

Our PICASO simulations show Bond albedos between 0-0.2 for hot clear Jupiters. Since the highest Bond albedo we found was near 0.2, we expect that any albedos observed to be greater than this must be caused by atmospheric aerosols, rather than Rayleigh scattering. Furthermore, we found that Bond albedos decrease with increasing metallicities, as the higher metallicity atmospheres will have greater abundances of opacity sources and so absorb more starlight. Planets with solar metallicities have Bond albedos of less than approximately 0.125, while planets with 100x solar metallicities have Bond albedos of approximately 0.025. Last, Bond albedo has little dependence on equilibrium temperature for planets less than 2000 K, and then decreases with increasing temperature.  We therefore choose a Bond albedo of 0.10 for the solar metallicity hot Jupiters discussed in this paper. \textcolor{black}{We chose this value to capture the contribution from Rayleigh scattering, and approximately match the values found from our grid of PICASO simulations, as shown in Figure \ref{fig:bond_albedo_grid}. We do not necessarily expect these values to exactly match the true values of HD 189733 b and HD 209458 b but rather to facilitate the radiative transfer investigations presented here. Optical eclipse measurements of HD 189733 b and HD 209458 b show geometric albedos of 0.076$\pm$0.016 and 0.096$\pm$0.016 respectively \citep{Brandeker2022, Krenn2023}. Assuming Rayleigh scattering is the only source of scattering, this results in Bond albedos of approximately 0.13 \citep{Heng2021}}, so in fact our choices are in good general agreement, considering that the presence of clouds could introduce additional albedo on these planets. We tested how well a global Bond albedo performed when used with picket fence radiative transfer. Our new Rayleigh scattering scheme produces reasonable pressure temperature profiles when compared to more sophisticated codes over a range of planet characteristics.

\subsection{Model Parameters}
\null\newpage

\begin{table}[ht]
\centering
\begin{tabular}{lccc}
\toprule
\multicolumn{4}{c}{HD 209458 b} \\
\midrule
Radiative Transfer & Clouds Type & Condensation Fraction & Timesteps per day \\
\midrule
Double Gray & No clouds & N/A & 4800 \\
Double Gray & All species compact clouds & 0.1 & 4800 \\
Double Gray & Default compact clouds & 0.1 & 4800 \\
Double Gray & All species clouds & 0.025 & 9600 \\
Double Gray & Default clouds & 0.10 & 9600 \\
Picket Fence & No clouds & N/A & 4800 \\
Picket Fence & All species compact clouds & 0.1 & 4800 \\
Picket Fence & Default compact clouds & 0.1 & 4800 \\
Picket Fence & All species clouds & 0.025 & 9600 \\
Picket Fence & Default clouds & 0.025 & 9600 \\
\midrule
\multicolumn{4}{c}{HD 189733 b} \\
\midrule
Radiative Transfer & Cloud Type & Condensation Fraction & Timesteps per day \\
\midrule
Double Gray & No clouds & N/A & 4800 \\
Double Gray & All species compact clouds & 0.1 & 4800 \\
Double Gray & Default compact clouds & 0.1 & 4800 \\
Double Gray & All species clouds & 0.05 & 4800 \\
Double Gray & Default clouds & 0.05 & 9600 \\
Picket Fence & No clouds & N/A & 4800 \\
Picket Fence & All species compact clouds & 0.1 & 4800 \\
Picket Fence & Default compact clouds & 0.1 & 4800 \\
Picket Fence & All species clouds & 0.025 & 4800 \\
Picket Fence & Default clouds & 0.05 & 4800 \\
\bottomrule
\end{tabular}
\label{tab:all_model_parameters}
\caption{For some models we were forced to adjust our default parameters in order be computationally tractable. Here we give the condensation fractions ($f$) and time resolution (given as number of time steps within each planet day, i.e., one rotation period) used for each model.}
\end{table}

\null\newpage
%\bibliography{bib}{}
\bibliographystyle{aasjournal}

\end{document}